\documentclass[twocolumn,tighten,times]{aastex63}
\usepackage{amssymb}
\usepackage{amsmath}
\usepackage{amsfonts}
\usepackage{graphicx}
\usepackage{txfonts}
\usepackage{enumitem}
\usepackage{xspace}

\def\be{\begin{equation}}
\def\ee{\end{equation}}
\def\ba{\begin{eqnarray}}
\def\ea{\end{eqnarray}}

\newcommand{\DM}{\ensuremath{\mathrm{DM}}}
\newcommand{\DMhat}{\ensuremath{\widehat{\mathrm{DM}}}}
\newcommand{\SM}{\ensuremath{\mathrm{SM}}}
\newcommand{\Dtd}{\Delta t_{\mathrm{d}}}
\newcommand{\Dnud}{\Delta \nu_{\mathrm{d}}}
\newcommand{\taud}{\tau_{\mathrm{d}}}
\newcommand{\niss}{n_{\mathrm{ISS}}}

\newcommand{\cnsq}{C_n^2}
\newcommand{\PSR}{PSR~J2219$+$4754}
\newcommand{\fJ}{f_{\rm J}}

\newcommand{\Np}{N_{\rm p}}

\defcitealias{css16}{CSS}
\defcitealias{Donner+19}{DVT+}
\defcitealias{Michilli+18}{MHD+}

\newcommand{\DV}{\citetalias{Donner+19}\xspace}
\newcommand{\MH}{\citetalias{Michilli+18}\xspace}

\renewcommand{\DV}{\citet{Donner+19}\xspace}
\renewcommand{\MH}{\citet{Michilli+18}\xspace}

\renewcommand{\added}[1]{{#1}}

\shorttitle{The Frequency-Dependent DM of PSR~J2219+4754}
\shortauthors{Lam et al.}

\begin{document} 
\title{On Frequency-Dependent Dispersion Measures and Extreme Scattering Events}

\author[0000-0003-0721-651X]{M.\,T.\,Lam}
\affiliation{School of Physics and Astronomy, Rochester Institute of Technology, Rochester, NY 14623, USA; mtlsps@rit.edu}
\affiliation{Laboratory for Multiwavelength Astronomy, Rochester Institute of Technology, Rochester, NY 14623, USA}
\affiliation{Department of Physics and Astronomy, West Virginia University, White Hall, Morgantown, WV 26506, USA}
\affiliation{Center for Gravitational Waves and Cosmology, West Virginia University, Chestnut Ridge Research Building, Morgantown, WV 26505}

\author{T.~J.~W.~Lazio}
\affiliation{Jet Propulsion Laboratory, California Institute of Technology, 4800 Oak Grove Drive, Pasadena, CA 91109, USA}

\author[0000-0001-8885-6388]{T.\,Dolch}
\affiliation{Department of Physics, Hillsdale College, 33 E. College Street, Hillsdale, Michigan 49242, USA}

\author[0000-0001-6607-3710]{M.\,L.\,Jones}
\affiliation{Department of Physics and Astronomy, West Virginia University, P.O. Box 6315, Morgantown, WV 26506, USA}
\affiliation{Center for Gravitational Waves and Cosmology, West Virginia University, Chestnut Ridge Research Building, Morgantown, WV 26505, USA}

\author[0000-0001-7697-7422]{M.\,A.\,McLaughlin}
\affiliation{Department of Physics and Astronomy, West Virginia University, P.O. Box 6315, Morgantown, WV 26506, USA}
\affiliation{Center for Gravitational Waves and Cosmology, West Virginia University, Chestnut Ridge Research Building, Morgantown, WV 26505, USA}

\author[0000-0002-1797-3277]{D.\,R.\,Stinebring}
\affiliation{Department of Physics and Astronomy, Oberlin College, Oberlin, OH 44074, USA}

\author[0000-0002-9507-6985]{M.~Surnis}
\affiliation{Department of Physics and Astronomy, West Virginia University, P.O. Box 6315, Morgantown, WV 26506, USA}
\affiliation{Center for Gravitational Waves and Cosmology, West Virginia University, Chestnut Ridge Research Building, Morgantown, WV 26505, USA}

\begin{abstract}

Radio emission propagating over an Earth-pulsar line of sight provides a unique probe of the intervening ionized interstellar medium (ISM). Variations in the integrated electron column density along this line of sight, or dispersion measure (DM), have been observed since shortly after the discovery of pulsars. As early as 2006, frequency-dependent dispersion measures have been observed and attributed to several possible causes. Ray-path averaging over different effective light-cone volumes through the turbulent ISM contributes to this effect as will DM misestimation due to radio propagation across compact lensing structures such as those caused by ``extreme scattering events''. We present methods to assess the variations in frequency-dependent dispersion measures due to the turbulent ISM versus these compact lensing structures along the line of sight. We analyze recent Low-Frequency Array (LOFAR) observations of \PSR\ to test the underlying physical mechanism of the observed frequency-dependent DM. Previous analyses have indicated the presence of strong lensing due to compact overdensities halfway between the Earth and pulsar. Instead we find the frequency dependence of the DM timeseries for \PSR\ is consistent with being due solely to ISM turbulence and there is no evidence for any extreme scattering event or small-scale lensing structure. The data show possible deviations from a uniform turbulent medium, suggesting that there may be an enhanced scattering screen near one of the two ends of the line of sight. We present this analysis as an example of the power of low-frequency observations to distinguish the underlying mechanisms in frequency-dependent propagation effects.



\end{abstract}

   \keywords{ISM: structure --
   pulsars: individual: PSR~J2219$+$4754}

\section{Introduction}\label{sec:intro}

The propagation of radio pulsar emission through the interstellar medium (ISM) provides a unique probe of ionized material along the path of propagation \citep{Rickett90}. Estimates of the dispersion measure (DM), or integrated electron column density, are obtained from precisely measuring the arrival times of pulse emission as a function of radio frequency \citep{Stairs2002}. Measurements of how these dispersive delays, along with other frequency-dependent delays, change over time allow us to build models of structures in the ISM on a wide range of spatial scales \citep{Armstrong+95}.

While temporal DM variations have been observed for decades \citep{rr71,pw91,pw92,Backer+93,Hobbs+2004}, frequency-dependent DM due to differing volumes of the ISM probed has only been observed more recently (\added{see e.g., for PSR~B1937+21 in} \citealt{Ramachandran+2006}, \citealt{DemorestThesis}; see also in \citealt{PennucciThesis}). A thorough treatment on the theory of frequency-dependent DM was given by \citet{css16}. Ray-path averaging of the radio emission through different volumes, and therefore different electron content, of the turbulent ISM will result in a smoothing of the DM timeseries by a kernel that broadens rapidly at lower frequencies. 

Besides dispersive delays, there are a number of other physical mechanisms that cause the propagating radio emission to be delayed. These delays vary as a function of frequency \citep{fc90,Clegg+98,sc17}. One example is from lensing of the emission around compact over- or under-densities in the ISM \citep{Clegg+98,Cordes+17}. However, since typical pulsar timing models only account for dispersive delays, there will be errors associated with the estimated DM due to these additional delays \citep{cs10,DMt}. Therefore, DM estimates taken at two separate frequencies may show differences solely due to biases from these other propagation effects and can be mis-attributed to a frequency dependence in the true dispersive delay.

\citet{Donner+19} have most recently reported three-and-a-half years of timing measurements of \PSR\ (B2217$+$47) with three stations of the International LOFAR  (LOw-Frequency ARray) Telescope (ILT). From those data, they have determined frequency-dependent variations in the \DM. The temporal variations noted in the DM timeseries for \PSR\ for both the lower- and higher-frequency data qualitatively appear very similar to those predicted by \citet{css16} for a turbulent medium, especially in that the higher radio-frequency DM data show higher fluctuation/Fourier-frequency structure than the lower radio-frequency timeseries. \DV\ concluded, however, that while the data are consistent with arising from the turbulent medium, the DM variations result from small-scale ($\sim$1~AU) structure(s) in the ISM, potentially multiple extreme scattering events (ESEs).

ESEs were originally seen as localized ``events'' in flux density measurements of compact radio sources \citep{Fiedler+87,Cognard+93,mlc03}. The first measurements of ESEs observed in DM and scintillation/scattering measurements were performed by \cite{Coles+15}, who described ESEs observed towards two pulsars. 

\added{This line of sight is particularly unique due to observed ``light echoes'' in the trailing components of the pulsar's pulse profile \citep{Michilli+18} using LOFAR. These additional components were seen to vary with time. \MH described several possible mechanisms, with one requiring the lensing of radio emission due to  passage near a compact interstellar structure, like those causing ESEs. Such lensing would cause the emission to traverse slightly different paths through the ISM, which could manifest as some combination of dispersive delays and light-travel-time delays. Given the complexity of the observed features, the line of sight toward \PSR\ therefore represents an excellent laboratory to test the predictions of \citet{css16} to disentangle the properties of the intervening medium. }


In this work, using the data on \PSR\ as a test to distinguish between the different physical mechanisms, we carry out analyses to determine whether or not the observations are consistent with lensing events or with the theoretical expectation from a turbulent medium. In \S\ref{sec:sf}, we build upon their structure-function analysis to show consistency of the frequency-dependent DM with the theoretical predictions for a Kolmogorov medium in \citet{css16}. In \S\ref{sec:ese}, we further examine 
the reported ESE detection, analyzing the impact of a variety of timing effects on their results. \added{We apply a wide range of analyses discussed in the literature to understanding this specific line of sight, including:} (i) determining the probability for the false signature of a gradient to be seen in the DM timeseries; (ii) establishing the importance of the solar wind on the frequency-dependence of DM and testing if a DM gradient is due to the changing Earth-pulsar line of sight; (iii) determining the impact of non-dispersive delays on DM estimation, specifically the amplitude of the delay caused by refraction from a lensing structure; (iv) understanding the systematic bias of pulse profile evolution in both time and frequency on these analyses; (v) accounting for various components of pulse arrival-time uncertainties, demonstrating the importance of high-precision pulsar timing techniques given the sensitivity of the measurements; (vi) developing a method built upon traditional structure function analyses, providing a more robust measure of the expected amplitude of the frequency-dependent DM while further showing inconsistencies with gradients in the DM timeseries; and (vii) implementing this structure function method on the data presented in \DV to gain additional insights on the line of sight. We found through our analyses that any lensing structure as described by \DV would affect the various observables in a way that has not been seen, thereby ruling out such a structure. We discuss the impact on precision pulsar-timing experiments in \S\ref{sec:timing} and briefly discuss possible future observations of this type in \S\ref{sec:discussion}. In the Appendix, we provide a derivation for DM estimation errors in the presence of time-of-arrival (TOA) uncertainties and additional frequency-dependent time delays that is useful for a number of arguments in our work.



\section{Structure Function Analysis and Frequency-dependent DM}\label{sec:sf}

Structure functions are used in analyses of pulsar DM timeseries, as well as the broader literature on turbulence in general, as a method of constraining the spectral properties of the variations \citep{cr98}. From the structure function, one can derive the amplitude of the electron-density wavenumber spectrum, which directly relates to the size of the fluctuations seen in DM timeseries \citep{Rickett90}. Therefore, they can also be used to quantify the magnitude of the differences between DM timeseries measured at two frequencies \citep{css16}. The variations in the observations presented in \DV provide a strong data set with which to test for the root-mean-square (rms) fluctuations in the DM differences with both frequency and time.

As part of their analysis, \citet[][see their Figure 6]{Donner+19} determined the structure function of the $\DM(t)$ timeseries, defined as $D_\DM(\tau) = \left<\left[\DM(t+\tau) - \DM(t)\right]^2\right>$, where $\tau$ is the time lag separating two observations and the brackets denote the ensemble average.  
The authors found a power-law structure function with a spectral index consistent with that of Kolmogorov turbulence.
The fit was performed over lags $\tau \le 200$~days. \added{Note that} the calculated structure function shows clear evidence of a white noise contribution at lags $\tau \lesssim 30$~days.  A white noise contribution to a time series will produce a ``plateau'' at small lags \citep[e.g.,][]{cd85} or for any trends (e.g., linear) in the data \citep{DMt,NG9DM} and bias the fit for the spectral index which will itself become a function of $\tau$. 
For the purposes of this paper, we accept that a Kolmogorov (or near-Kolmogorov) spectrum is an acceptable fit to the DM structure function given the clear overlap with the estimated values shown. We will assume from here that the spectrum is described by a Kolmogorov medium based on the observed consistency with this model in the ISM \citep[e.g.,][]{Armstrong+95}.

Following the determination of a Kolmogorov spectrum, the authors then concluded that the cause of the variations may be due to ESEs, which they describe in a context other than localized events. A spectrum steeper than Kolmogorov or discrete structures (or both) are required to produce ESEs and the refractive effects observed in pulsar dynamic spectra  \citep[e.g.,][]{ra82,hwg85,cw86,rbc87}. The turbulence in the ISM is often described with an electron-density wavenumber spectrum having a power-law form \citep[see e.g.,][]{Armstrong+95}
\be 
P_{\delta n_{\mathrm{e}}} = C_n^2 q^{-\beta}, \quad q_1 \le q \le q_2
\ee
where $q = 2\pi/l$ is the wavenumber corresponding to physical scale $l$ and $C_n^2$ is the spectral coefficient representing the overall amplitude or strength of the turbulence. The low and high wavenumber cutoffs represent outer and inner physical scales, respectively. A Kolmogorov medium corresponds to $\beta = 11/3$. Strong refractive effects occur for a density spectral index $\beta > 4$ \citep{cpl86}, which was not observed for \PSR. Notably, \DV\ referenced \cite{Fiedler+94} and \cite{Coles+15} as examples of ESEs in which larger-scale turbulent structure is identified as the cause of the observed lensing; however, both of those works instead find the causes of their observations to be due to structures consistent with compact canonical ESEs, i.e., smaller-scale lenses as described previously in the literature and above. In the former, the authors did examine the associated large-scale structure of the ISM in the foreground of sources with observed ESEs from \cite{Fiedler+87}.  In the latter, the outer scale of the ESE was taken to be of order the size scale of the smallest dimension of the lens, which is distinct from the outer scale of the electron-density wavenumber spectrum of the ionized ISM written above, known to be many orders of magnitude larger \citep{Armstrong+95}.

Using the constant value of the amplitude of the wavenumber spectrum \DV estimated, $\cnsq = 0.9 \times 10^{-3}$~m$^{-20/3}$, and a distance estimate of 2.2~kpc from the NE2001 electron density model \citep{NE2001}, the corresponding scattering measure is $\SM = \int_0^{D_p} \cnsq(z) dz = 2.0 \times 10^{-3}$~kpc~m$^{-20/3}$, an order of magnitude higher than the value predicted by the NE2001 model\footnote{We note this measurement (and extrapolated to the estimate of the scintillation timescale discussed later) may then be useful in constraining properties in future electron density models or uncovering interesting turbulence physics along this line of sight.}.  Following \citet[][see Eq.~12]{css16}, the rms difference in \DM\ between two spot frequencies $\nu_1$ and $\nu_2$, where $\nu_1 < \nu_2$, for a uniform Kolmogorov medium is
\ba
\sigma_{\DM(\nu)}(\nu_1,\nu_2) & = & 
3.76 \times 10^{-5}~\mathrm{pc~cm}^{-3}~F_{11/3}(r)~\left(\frac{G_{11/3}}{0.145}\right)\left(\frac{D_p}{\mathrm{1\,kpc}}\right)^{5/6} \nonumber\\
& & \times  \left(\frac{\nu_2}{\mathrm{1\,GHz}}\right)^{-11/6}~\left(\frac{\SM}{10^{-3.5}~\mathrm{kpc~m}^{-20/3}}\right), 
\label{eq:sigmaDMnu_uniform}
\ea
where the frequency ratio $r \equiv \nu_2/\nu_1$, $F_{11/3}(r)$ is a factor that contains all of the relative frequency dependence \citep[][see Eq.~11]{css16}, $G_{11/3}$ is a geometric-only factor where the fiducial value of 0.145 is for a uniform Kolmogorov medium \citep[][see Table~1]{css16}, and $D_p$ is the pulsar distance.  For frequencies at the centers of the two frequency bands from the observations of \DV, we have $\sigma_{\DM(\nu)}(133~\mathrm{MHz},169~\mathrm{MHz}) = 0.6 \times 10^{-3}$~pc~cm$^{-3}$, while for the extremes of the bands, $\sigma_{\DM(\nu)}(118~\mathrm{MHz},190~\mathrm{MHz}) = 1.1 \times 10^{-3}$~pc~cm$^{-3}$. The bottom panel of Figure~5 of \DV shows rms variations at exactly these levels, suggesting that the observed frequency-dependent DM is consistent with the theoretical predictions of \citet{css16} for a Kolmogorov medium.


\section{Testing the ESE Interpretation}\label{sec:ese}

Frequency-dependent DM results from raypath averaging over different volumes of the intervening medium. Refraction by discrete structures along the propagation path can also result in variations in DM timeseries via mis-estimation from non-dispersive delays \citep{fc90,Cordes+17}. These delays can therefore be useful in characterizing such discrete structures \citep[e.g.,][]{secondISMevent}. However, care needs to be taken in order to properly account for all effects in timing data.

Following their analysis of the structure function, \DV proposed an interpretation of the DM variability due to plasma lensing structures. For simplicity, they made an assumption that the variability is due to one to three spherical lenses. Under this assumption, they found the potential parameters of one of the clouds (see their Figure 7). In this section, we will present an alternative explanation from a lens as described by \DV via characterization of the pulse arrival times in conjunction with the DM timeseries.


Qualitatively, the appearance of the DM time series is similar to that of other pulsars; for example, in the DM time series presented by \citet{NG9DM}, one can identify similar features for PSR~J0613$-$0200, PSR~B1937+21, and potentially other pulsars in that dataset alone.

Evidence of these features being caused by something other than ESEs can be found by considering the expected DM amplitude.  \DV\ identified the variation in DM between MJD~56950 and~57100 (their Figure~1) as an ESE by virtue of its amplitude $|\delta\DM| \approx 3 \times 10^{-3}$~pc~cm$^{-3}$.  By considering their calculated structure function and relating that to the expected rms variations in the \hbox{DM} \cite[Eq.~30]{DMt}, we find
\begin{eqnarray}
\sigma_{\DM}(\tau)
 &=& \left[\frac{1}{2} D_\DM(\tau)\right]^{1/2} \nonumber \\
 &=&  1.2 \times 10^{-5}~\mathrm{pc~cm}^{-3} \left(\frac{\tau}{\mathrm{day}}\right)^{5/6}.
\end{eqnarray}
For $\tau = 150$~days, then $\sigma_\DM(150~\mathrm{days}) = 0.8 \times 10^{-3}$~pc~cm$^{-3}$, and $|\delta \DM|$ is much larger than the expected rms variation.

If the DM variation is from a true steep gradient, then we can use an estimate of the rms DM gradient \citep[rather than the rms DM,][]{DMt},
\be 
\sigma_{d\DM/dt} \approx \frac{\sigma_{\DM}(\tau)}{\tau},
\ee
and compute the ``signal-to-noise'' ratio for such a gradient,
\be 
R_{d\DM/dt} \equiv \frac{\left|d\DM/dt\right|}{\sigma_{d\DM/dt}}.
\ee
Evaluating for $|d\DM/dt| = 2 \times 10^{-5}$~pc~cm$^{-3}$~day$^{-1}$, we find that $R_{d\DM/dt} = 3.8$ over the 150-day duration of the purported ESE, suggesting moderate significance. However, while the DM obtained from analyzing the entire frequency range shows a rising linear trend prior to this time period, in the split-frequency DM timeseries, the $<149$~MHz values are consistent with a constant DM within $\sim 1\sigma$, and therefore there is no apparent ingress into the assumed spherically-symmetric lensing material for those frequencies. Therefore, a canonical ESE interpretation seems unlikely, though a potential DM gradient may still be plausible. Note that the egress time period occurs around MJD 57100, right when the DM difference between the timeseries is zero by construction. It is thus unclear if this time period truly marks the end of such an event or not.

\begin{figure}[t!]
\centering
\includegraphics[width=0.5\textwidth]{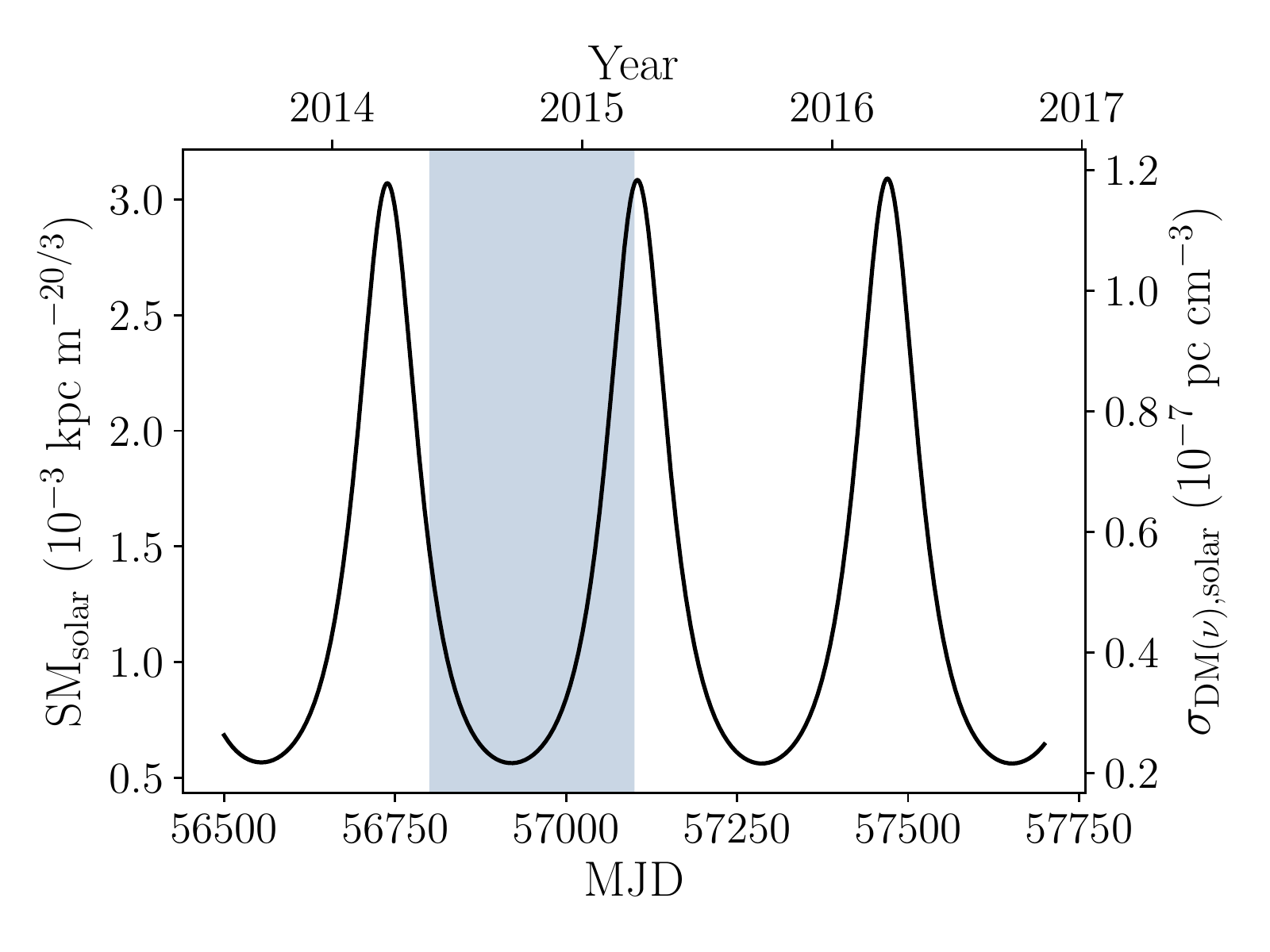}
  \caption{\footnotesize The SM due to the solar wind along a line integrating out from the Earth to the direction of \PSR. The right axis shows the equivalent $\sigma_{\DM(\nu)}$ uncertainty. The blue shaded region denotes the 300-day timespan of the proposed lens.
  }
\label{fig:SMsolar}
\end{figure}

We investigated whether this potential DM gradient could be due to the line of sight cutting through different parts of the solar wind. By MJD 56950 moving forward, the suggested time of a peak, the solar elongation is shrinking (for reference, the pulsar's ecliptic latitude is 52.5$^\circ$). Thus, as the pulsar as seen on the sky approaches the Sun we would expect the DM to increase (only slightly due to the high ecliptic latitude) rather than decrease \citep{You+07b,NG9DM,NG11SW}. Any solar flare or coronal events that may have occurred during this timespan would also cause an increase rather than a decrease in the DM \citep{DMt}. In addition, we looked at the $C_n^2$ contribution due to turbulence in the solar wind using the form provided in \cite{Spangler+02},
\be 
C_n^2(r) = 1.8 \times 10^{10}~\left(\frac{r}{10~R_\circ}\right)^{-3.66}~\mathrm{m}^{-20/3},
\ee
where $r$ is the radial distance from the Sun. We integrated the solar wind $C_n^2$ over the line of sight to determine the SM, shown in Figure~\ref{fig:SMsolar}. With each integration element acting as a thin screen, we determined the frequency-dependent DM error and then combined these to find a total error of
\be 
\sigma_{\DM(\nu),\mathrm{solar}} =  \sqrt{\frac{1}{\left< 1/\sigma_{\DM(\nu)}^2(s)\right>}},
\ee
i.e., the total error is the square root of the reciprocal of the mean of the ``weights'' ($1/\sigma^2$), and we are integrating along the line-of-sight position $s$ for each volume element. We find that while the SM is increasing over the time of the suggested ESE and is of a similar value to the rest of the ISM (fiducial value of 10$^{-3.5}$; \citealt{Rickett90}, \citealt{cr98})---because the material is very close to the Earth in comparison with the pulsar's distance---the rms frequency-dependent DM is small, well below the measurement uncertainties and even below that of other high-precision timing experiments \citep[see e.g.,][]{NG9DM}.

For variations in DM due to the changing ionosphere, in this case because of the yearly modulation from observing the pulsar transitioning between day and night over that time period, the amplitude of the change in DM is at most $\sim$10$^{-4}$~pc~cm$^{-3}$ and therefore is not a significant contributor to the variation here \citep{DMt}.

\subsection{Impact of Non-Dispersive Delays}

As has been described in the literature, there are multiple frequency-dependent delays that affect pulsar TOAs \citep{fc90,DMt,secondISMevent,sc17}.  Beyond the traditional dispersive delay due to the integrated electron density ($\propto\nu^{-2}$), there is a geometric delay ($\propto\nu^{-4}$) due to total path length changes and a  barycentric-correction delay ($\propto\nu^{-2}$) due to the angle of arrival of the pulsar shifting, i.e., the pulsar's image appearing from a different direction on the sky.

Following the notation in \cite{secondISMevent}, the dispersive delay for a lens of size $L$, electron density $n_e$, and dispersion measure $\DM_l \sim n_e L$, is 
\be 
t_\DM \sim \frac{\lambda^2 r_e \DM_l}{2 \pi c} \sim  \frac{\lambda^2 r_e n_e L}{2 \pi c},
\ee
where $\lambda$ is the electromagnetic wavelength, $r_e$ is the classical electron radius, and $c$ is the speed of light. The geometric delay is 
\be 
t_{\rm geo} \sim \frac{D_l (1-D_l/D_p) \lambda^4 r_e^2 (\DM_l^\prime)^2}{8 \pi^2 c} \sim \frac{D_l (1-D_l/D_p) \lambda^4 r_e^2 n_e^2}{8 \pi^2 c \zeta^2},
\ee
where $D_l$ and $D_p$ are the distance to the lens and pulsar (from the observer), respectively, and $\DM_l^\prime \sim n_e L / (\zeta L) \sim n_e/\zeta$ is the DM spatial gradient with the depth-to-length aspect ratio of the lens as $\zeta$. Finally, the barycentric delay is
\be 
t_{\rm bary} \sim \frac{(1 - D_l/D_p)\lambda^2 r_\oplus r_e \DM_l^\prime}{2\pi c} \sim \frac{(1 - D_l/D_p)\lambda^2 r_\oplus r_e n_e}{2\pi c \zeta},
\ee
where $r_\oplus$ is the Earth-Sun distance of 1~AU.

\begin{figure}[t!]
\centering
\includegraphics[width=0.5\textwidth]{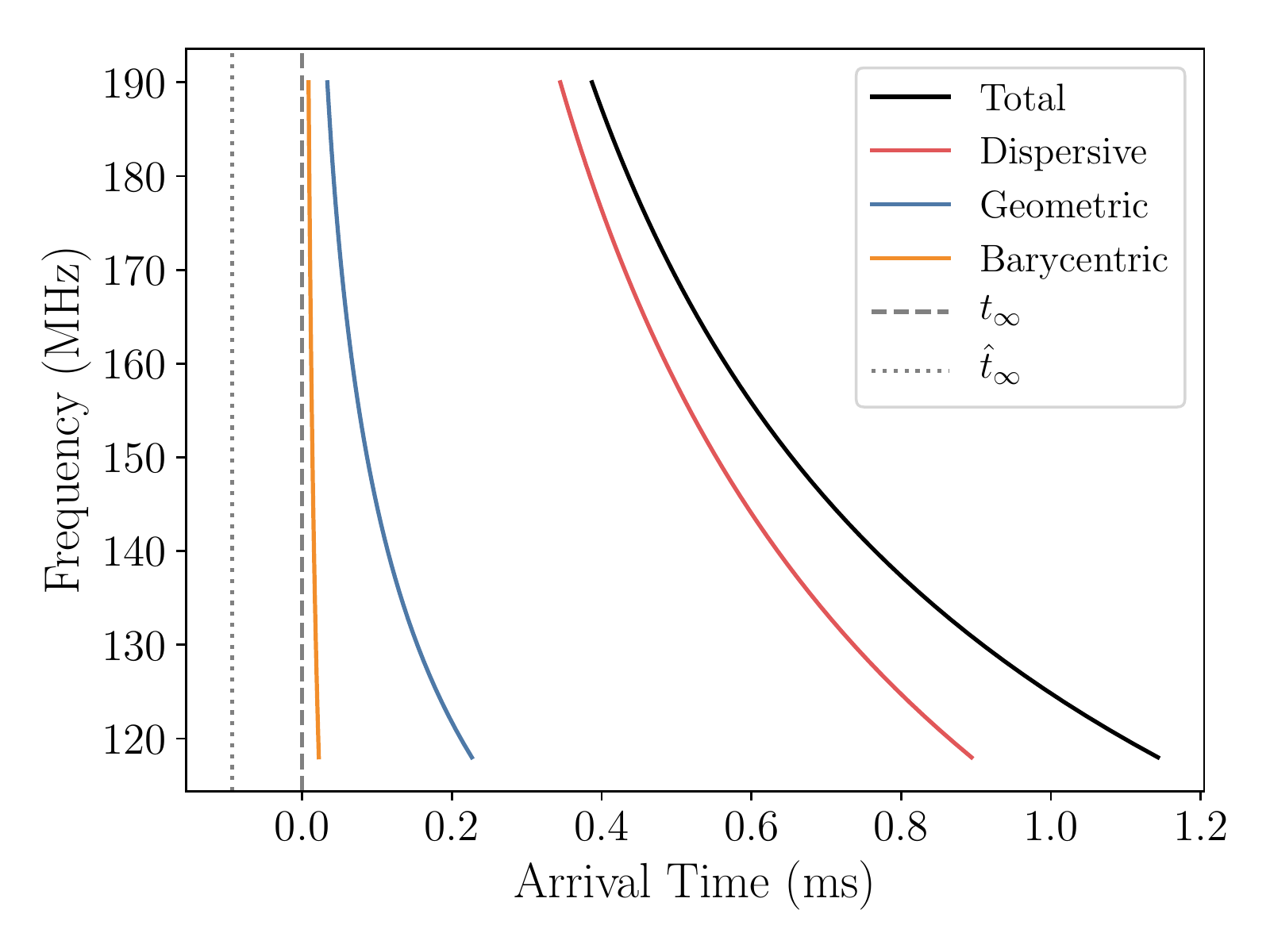}
  \caption{\footnotesize Pulse delays as a function of time and frequency. The total delay is the sum of the dispersive, geometric, and barycentric delays. The dashed gray line shows the true infinite-frequency arrival time (set to zero) while the dotted gray line shows the estimated infinite-frequency arrival time when a purely dispersive delay is fit over all frequencies to the total delay (i.e., extrapolating the delay curve to infinite frequency), $\delta t_\infty = -93~\mu$s. We assume a cloud 1.1~kpc away from the Earth, the estimate for the distance in \MH, which implies a size $L = 20$~AU and central density $n_e = 31$~cm$^{-3}$ (from our analysis but also see Figure 7 of \DV). For clarity, the legend from top to bottom displays the different curves from right to left.
 }
\label{fig:threedelays}
\end{figure}

\begin{figure}[t!]
\centering
\includegraphics[width=0.5\textwidth]{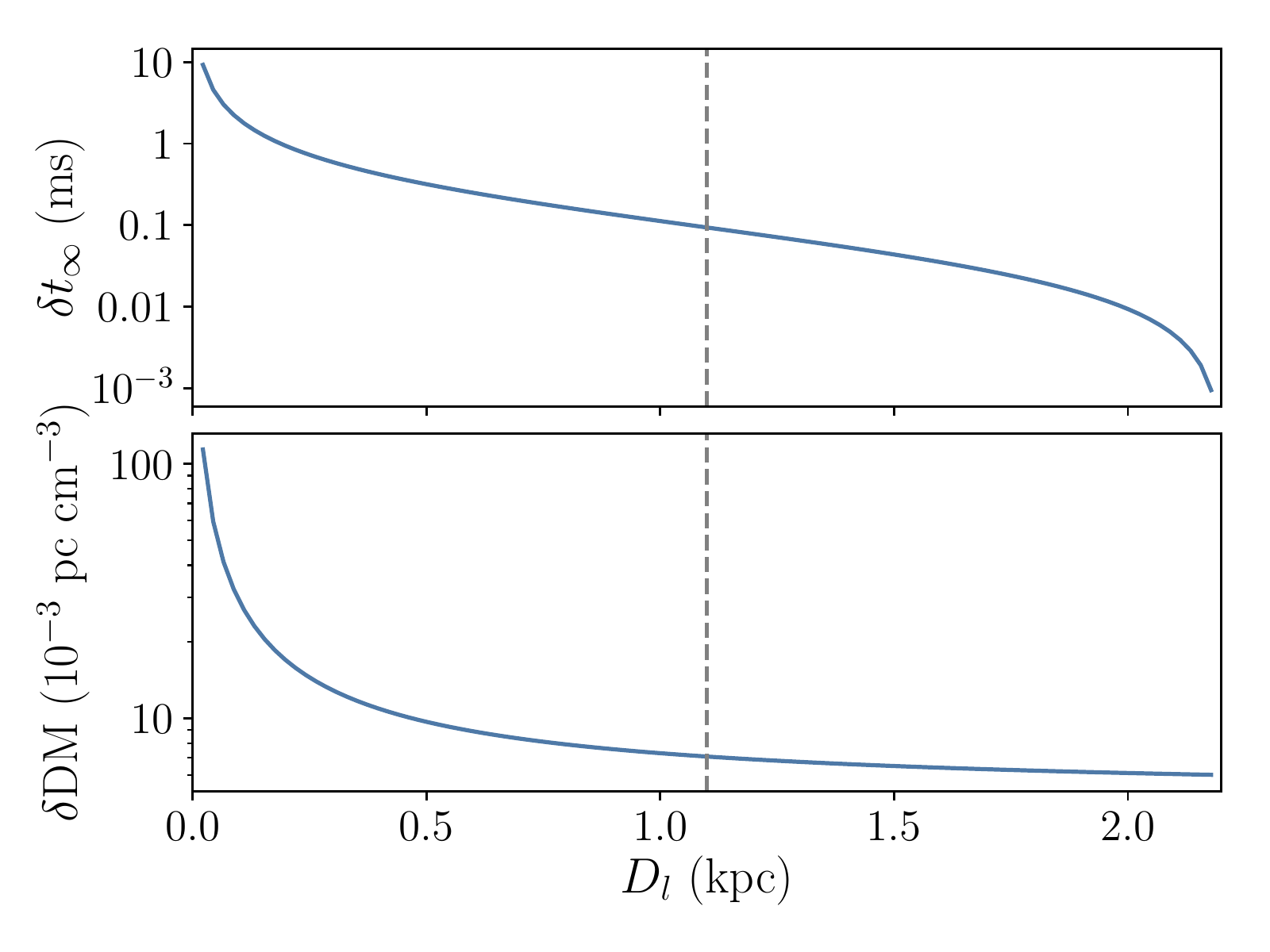}
  \caption{\footnotesize The TOA perturbation (top) and DM perturbation (bottom) when fitting for the set of total frequency-dependent delays (e.g., see Figure~\ref{fig:threedelays}) versus only the dispersive delay. We display these perturbations as a function of the lens distance.}
\label{fig:lens_perturbations}
\end{figure}

Following \DV, we assume that the cloud is spherical and therefore we set $\zeta = 1$. The time for the cloud to pass through the line of sight is $\sim 300$~days. With a proper motion of 22.2 mas/yr \citep{Michilli+18}, the angle subtended on the sky is $\theta = 18.2$~mas. The angle sets the physical size of the cloud as $L = \theta D_l$. Therefore, the distance of the lens is linearly proportional to the size of the lens \added{(shown by \citealt{Donner+19} in the bottom panel of their Figure 7)}. Given that the DM change from the cloud is simply $n_e L$, we have that $n_e$ is inversely proportional to the cloud size and inversely proportional to the lens distance \added{(shown by \citealt{Donner+19} in the top panel of their Figure 7)}.

The distance estimate for a possible structure in the ISM causing light ``echoes'' seen in the pulse profiles for \PSR\  is $D_l \approx 1.1$~kpc \citep[][hereafter \MH]{Michilli+18}, a companion paper to the work of \DV. Using the DM change of $\left|\delta \DM\right| \approx 3 \times 10^{-3}$~pc~cm$^{-3}$, Figure~\ref{fig:threedelays} shows the dispersive, geometric, and barycentric delays for such a lensing structure at 1.1~kpc. We see that the dispersive delay dominates over the geometric and barycentric delay. However, when fitting the total frequency-dependent delays observed, the estimated infinite-frequency arrival time is shifted from the true arrival time by $93~\mu$s; the estimated DM would then be larger than $\left|\delta \DM\right|$ above by $7.1 \times 10^{-3}$~pc~cm$^{-3}$, which is unseen in the timeseries unless the baseline DM value is far lower than suggested by \DV.

In Figure~\ref{fig:lens_perturbations}, we show the perturbation in $t_\infty$ and DM (the difference between the ``true'' DM from the dispersive delay and that from the mis-estimation) after fitting the total delay curve generated by placing a lens at a distance $D_l$; again, Figure~\ref{fig:threedelays} shows the three frequency-dependent delays along with the total delays when $D_l = 1.1$~kpc. These perturbations were calculated after fitting each delay curve with the functional form $t_\nu = t_\infty + K\DM/\nu^2$. For reference, the form of these perturbations for two spot frequencies are shown in the Appendix.

\begin{figure}[t!]
\hspace{-3ex}
\includegraphics[width=0.525\textwidth]{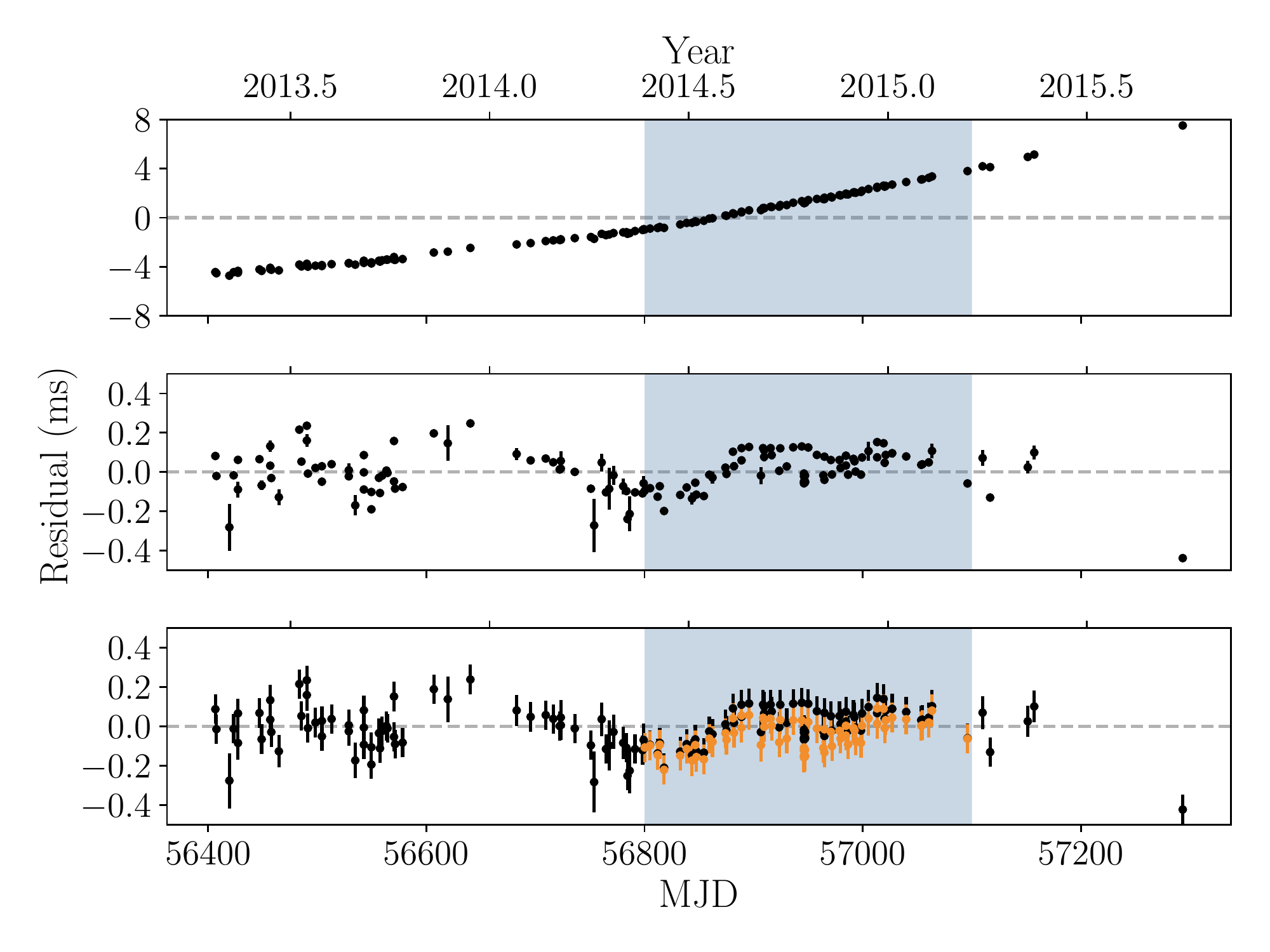}
  \caption{\footnotesize Timing residuals from \MH. Top: Original residual values as plotted in \MH, after a fit for spin and astrometric parameters, a mean DM, and one glitch. The steep trend is likely due to longer-term spin noise \citep{sc10}. Middle: For visual clarity only on the scatter of the residuals, we have removed the best-fit quadratic trend. Bottom: In black are the same residuals as shown in the middle panel but with the contributions of jitter ($\approx 10~\mu$s) and polarization miscalibration ($\approx 80~\mu$s) added in quadrature. The quadratic trend was not refitted since it is only used as a visual aid. The blue shaded regions in all panels denote the 300-day timespan of the proposed lens. The red points in the bottom panel show the impact of a lens with timing perturbations to $t_\infty$ represented as a triangle function with the amplitude going from 0 to $-93~\mu$s in the middle and back to 0.
  }
\label{fig:residuals}
\end{figure}

If we split the total band into two halves to measure the frequency-dependence of DM, the increases in $\delta \DM$ are then 7.3 and $6.8 \times 10^{-3}$~pc~cm$^{-3}$ for the low and high bands, respectively, or a difference of $5 \times 10^{-4}$~pc~cm$^{-3}$ between the two. This value is approximately half of the difference between the two bands shown by \DV in their Figure~5. Recall, however, that in this calculation we started with a single DM, and only when estimating the DM in both halves of the band did we recover frequency dependence, solely from the unaccounted for $\propto\nu^{-4}$ geometric delay rather than from a true DM difference. This 
demonstrates why a complete analysis of the arrival times, such as in \cite{secondISMevent}, is crucial to understanding any potential lens/ESEs in the data rather than analyzing only the DM timeseries.

\added{We examined the} long-term \added{timing} residuals in \citet[][see their Figure 1]{Michilli+18} for \PSR\ from observations covering 1970 to 2016. Figure~\ref{fig:lens_perturbations} shows that the TOA perturbations due to a lens should be of order microseconds or greater depending on the distance, and for a lens at distance 1.1~kpc, we expect a perturbation of $\approx 0.1$~ms if $\propto \nu^{-4}$ delays are not fit for. We replot the residuals \added{as directly shown in} \MH in Figure~\ref{fig:residuals}, starting with the earliest observation epoch in \DV. 
For reference, decimal year 2015.5 corresponds to the end of the time period in which \DV\ stated no baseline DM variations were observed. The blue shaded region denotes the 300-day time period over which the proposed lens occurs. In the middle panel, we show the residuals after a quadratic subtraction so that the scatter is more clearly visible. In the bottom panel, we show the same residuals as in the middle but with an increase in the TOA uncertainties due to jitter ($\approx 10~\mu$s) and polarization miscalibration ($\approx 80~\mu$s) as described later in  \S\ref{sec:TOAerrs}. In red we have injected the systematic effect on the timing given the lens described above, where we simplify input of the offsets as a triangle function starting at 0, decreasing to $-93~\mu$s in the middle, and increasing back to 0. Given the scatter in the residuals and the unknown contributions to spin noise, we cannot prove or disprove the presence of a lens at 1.1~kpc.  As per Figure~\ref{fig:lens_perturbations}, a lens closer than 1.1~kpc will show increasingly dramatic dips in the residual timeseries, which would become readily visible.


\MH also showed the evolution of the pulsar's spindown rate. While this curve is also smoothly varying, we see a drop right at the start of 2015 (around MJD 57000), during the time of the proposed ESE. \DV\ indicated that a single spin-period derivative (\MH\ showed the equivalent spin-frequency derivative) was fit in their own timing model. The spindown rate shown in \MH was not constant over this time period, suggesting that if the spin-frequency derivative was smaller than the average value in the fit by \DV, then the pulsar's spin was braking more rapidly (i.e. the spin frequency was dropping more rapidly) than expected and the infinite-frequency arrival times should be delayed compared to their model (a linear change would have been absorbed by the shorter-duration fit but we do not see that either). Any fit for DM would therefore be biased by this effect. Visualized in terms of the delay curves in Figure~\ref{fig:threedelays}, the true arrival time would be delayed, i.e., shifted to the right.

Rather than investigating the impact on the timing residuals, we directly investigated the impact on the two DM timeseries. Given the arguments above, if there was an intervening lens with a true column density equal to $\DM \approx 3 \times 10^{-3}$~pc~cm$^{-3}$ that passed the line of sight, the differences in DM between the two bands could potentially be explained, but not the amplitude change of $\approx 7 \times 10^{-3}$~pc~cm$^{-3}$ in the DM timeseries from the zero value, thereby ruling out a lens with such a column density. It is possible that if a lens with a lower column density were to pass by the line of sight, it would show up with an apparent DM having the appropriate amplitude as estimated by the total time delays. One can search over the phase space of the true column density/DM, size $L$, and distance $D_l$ to find the best-fit parameters of a possible lens. As stated previously, for a spherically symmetric lens, we would expect the DM timeseries from both bands to behave symmetrically.


In addition to the three delays described above, general scattering (alternatively, pulse broadening) from a Kolmogorov medium will produce arrival-time delays $\propto \nu^{-4.4}$. \DV\ stated that if the frequency-dependent DM is variable in time, then non-$\nu^{-2}$ dispersive delays will vary equally in time. From \citet{css16}, we see that the time- and frequency-dependence of DM arises naturally from ray-path averaging over different volumes of the ISM if there is an effective velocity between the Earth, pulsar, and bulk ISM motions \citep[regarding the effective velocity, see][]{cr98}. Such variations can arise even for a medium with constant $C_n^2$, which for example can yield a statistically constant scattering timescale even while the DM varies. Changes in pulsar scattering timescales have been observed \citep[e.g.,][]{Coles+15,Levin+16}, which require a change in $C_n^2$ \citep[or the inner scale of the turbulence, again see][]{cr98} and therefore the {\it statistics} of the DM variations, though the timeseries of each need not be one-to-one correlated. As with the geometric and barycentric delay, a $\propto\nu^{-4.4}$ scattering delay will also cause DM to be incorrectly estimated if not properly modeled. \DV noted that unmodeled scattering does affect their results but is small enough in amplitude that it does not change their conclusions substantially.

\subsection{Impact of Frequency- and Time-Dependent Profile Evolution}

\begin{figure}[t!]
\centering
\includegraphics[width=0.5\textwidth]{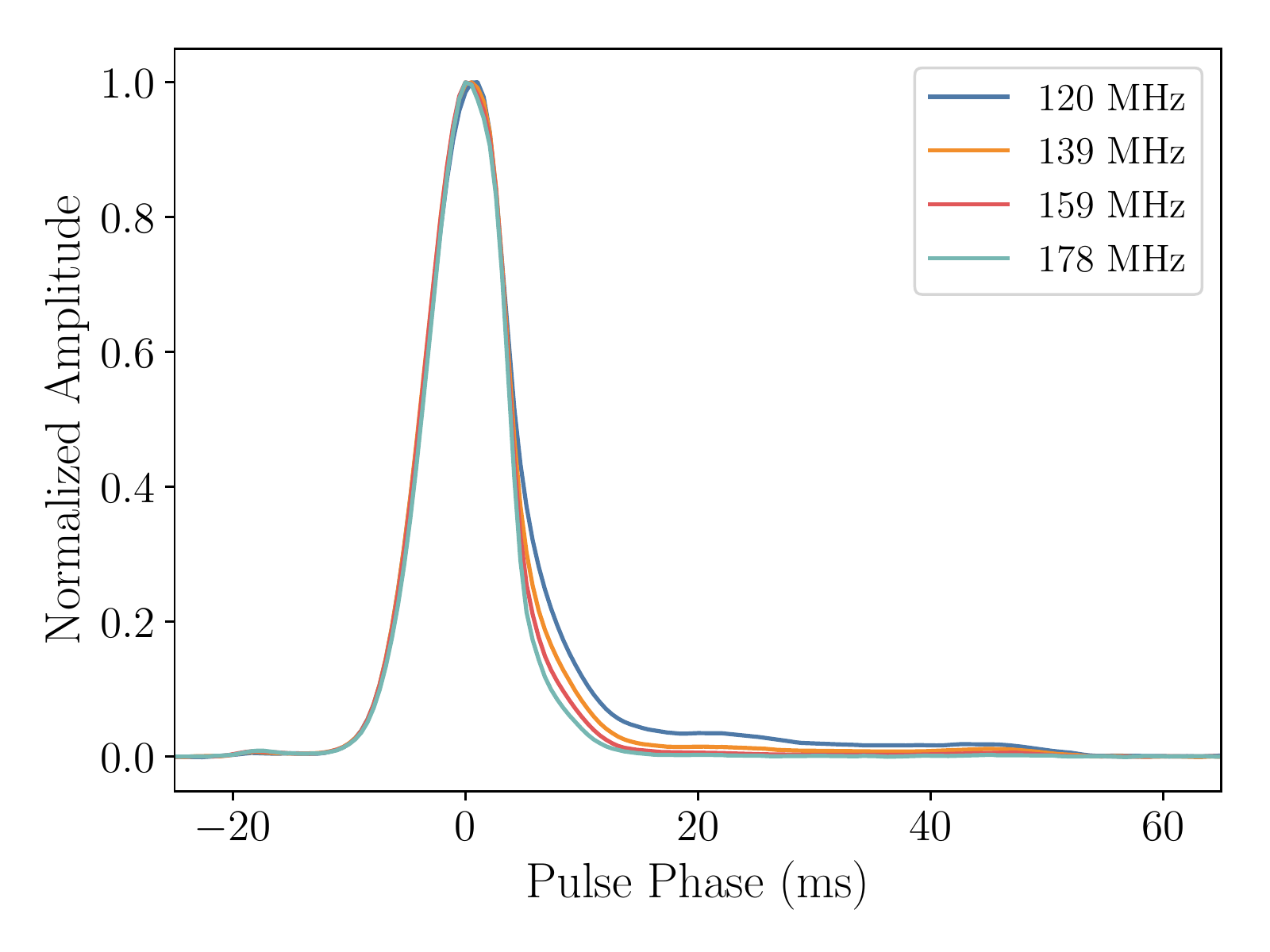}
  \caption{\footnotesize Profiles of \PSR\ as a function of frequency from \cite{Bilous+16}. The pulse phase is centered at the peak and only a small part around the main pulse is shown for clarity.}
\label{fig:profiles}
\end{figure}

The variability of the pulse profile of \PSR\ with time was discussed in \MH in the context of light echoes due to propagation effects. \cite{Bilous+16} showed in an observation taken in early 2014 (around MJD 56700) that there was significant frequency dependence in the pulse profile. Using profiles of \PSR\ from \cite{Bilous+16}\footnote{Obtained via the European Pulsar Network, \url{http://www.jb.man.ac.uk/research/pulsar/Resources/epn/}}, we examined the TOA and DM perturbations due to the frequency dependence of the profiles. These four profiles are shown in Figure~\ref{fig:profiles}; the trailing components are clearer at lower frequencies than at higher frequencies, suggesting that the shape variations as a function of time as seen in \MH are also a function of frequency, at least for some epochs.

\cite{Bilous+16} used their observations to determine a new spin period and DM for each of the pulsars in their census. The initial adjustment of these two parameters maximized the pulse signal-to-noise ratio. Afterward, an average template was generated, which was used to calculate more precise TOAs and perform a subsequent timing analysis to improve their period and DM estimate. They did not account for profile evolution, either intrinsic or from interstellar scattering, in their work but noted it as a potential bias; the profiles used in this analysis are therefore phase-aligned according to their method.

We generated smooth template shapes for each of the four profiles. While pulsar components are often fit with von Mises functions \citep[e.g.,][]{Oslowski+11,Hassall+12}, the circular analogue of a Gaussian function, the automated routine in the \textsc{PyPulse} package \citep{PyPulse} to fit multiple such components (see a more thorough description of the procedure conditions in \citealt{FDJitter}) did not converge for the lower-frequency profiles due to the shapes of the trailing components. We then used a simpler Savitzky-Golay filter \citep{SavitzkyGolay}, implemented in \textsc{scipy} with a moving 11-point cubic polynomial, which produced residuals between the template and the data profile at the level of the rms noise of the off pulse region; slight variations in the parameters did not affect the overall fit significantly. With our four smooth pulse shapes, we used \textsc{PyPulse} to fit the highest-frequency (178~MHz) template to the other three templates in order to compute the arrival-time offsets due to frequency-dependent profile evolution; these were 229.0, 75.7, and 11.2~$\mu$s for the 120, 139, and 159~MHz profiles, respectively. We chose the highest-frequency template as a standard of comparison to minimize the impact of the trailing component shapes as an approximation for the intrinsic pulse shape compared to a pulse shape showing a light echo.

Using these timing perturbations, we then calculated the estimated DM perturbations; see the Appendix for more on DM estimation in the presence of an additional chromatic perturbation, in this case delays due to the pulse shape changes as a function of frequency. Since the method of calculating DM between \cite{Bilous+16} and this analysis are different, we do not expect the absolute DM value to be the same \citep{DMt}. However, using the two lowest or highest frequencies to calculate the DM alone yielded a difference between the two of $1.7 \times 10^{-3}$~pc~cm$^{-3}$, again of the same amplitude of the variations seen by \DV. This value is an order of magnitude larger than the DM difference they quote between scattered and unscattered profiles. In performing their DM analysis, \DV accounted for frequency-dependent profile evolution with a frequency-resolved template, which should then mitigate any impact of this profile evolution on the measured DMs. As mentioned previously, since their template was generated from an observation on MJD 57161, we expect the DM at both bands to be equal by construction at that epoch, which is also noted by the authors, though the true absolute DM between the two at that time and over the course of the entire observing span may differ. Nevertheless, the changing profile shapes as a function of time as seen in \DV and in \MH could cause biases in the TOAs and the subsequent DM determinations, especially when coupled with the changes in frequency. Note again that \DV did state the amplitude of scattering in their data and the biases in their DM; \added{we find that} this amplitude \added{does not affect their conclusion that frequency-dependent DM is observed.} 

\subsection{Impact of Arrival-Time Uncertainties}\label{sec:TOAerrs}

We examined the role of additional components to the TOA uncertainties due to pulse jitter on the DM estimates. Traditionally until the last several years, many analyses considered the TOA uncertainty as arising only from template fitting \citep[e.g.,][]{Manchester+13,NG5,Desvignes+16}, the process of fitting a smoothed template shape to the data profile. The assumption of matched filtering which underlies this fitting is that the data is a shifted and scaled copy of the template shape. It has been long known that single pulses from pulsars vary stochastically \citep{CraftThesis}, implying that the data shape cannot be an exact copy of a template since the average of a finite number of single pulses will always be slightly different. In general, shape changes due to jitter include contributions both from phase and amplitude variations.

We used the jitter parameter $\fJ$ defined in \cite{cd85} for our analysis, defined as the ratio between the single-pulse rms jitter and the equivalent rms (i.e., for a Gaussian pulse, the standard deviation as compared to the full width at half maximum) of the template \citep{sod+14,NG9WN}. \cite{cs10} summarized a number of analyses in the literature, primarily for canonical pulsars, and suggest that $\fJ \approx 1/3-1/2$. \cite{sod+14} found similar values for millisecond pulsars. \cite{NG9WN,FDJitter} used a separate jitter parameter that is independent of the pulse shape but found comparable statistics for millisecond pulsars. In many cases, especially for bright canonical pulsars in which single pulses can be detected, jitter is the dominant component of the TOA uncertainty \citep{NG9WN}.

Again using profiles of \PSR\ from \cite{Bilous+16}, calculating the pulse width, and assuming a fiducial value of $\fJ = 1/3$, we found the rms single-pulse jitter for \PSR\ to be 1.2~ms, $\approx 0.2\%$ of pulse phase ($P = 0.5385$~s), consistent with some pulsars in \cite{NG9WN,FDJitter} though below the average. 
(Note that millisecond pulsar studies as described above were performed at significantly higher frequencies, and the statistics for millisecond pulsars may be different from that of canonical pulsars.) Since the rms jitter scales as the number of pulses $\Np^{-1/2}$, we assumed the pulsar was observed for the median observing time each epoch, 115 minutes, using the LOFAR station most used in \DV, DE605. Given the number of pulses in that time, we find that the TOA uncertainty due to jitter is $\approx 10~\mu$s. Using a larger value of the jitter parameter(s) will yield a larger uncertainty. In addition, changes in the integration time will vary the TOA uncertainty; the integration times for DE605 ranged from 2 to 146 minutes, which yield an equivalent rms jitter of 79~$\mu$s to 9~$\mu$s, respectively. 


We estimated the value of \added{the TOA} uncertainties as follows. \DV\ stated a median DM uncertainty of $3.7 \times 10^{-5}$~pc~cm$^{-3}$. Following the formalism of \cite{nonsimDM} and \citet{css16} of assuming that the (frequency-independent) DM is measured at two spot frequencies and then the infinite-frequency TOA is estimated by removing the dispersive delay, one can calculate the DM difference between the true DM and the estimated DM as 
\be 
\delta\DM = - \frac{\epsilon_{\nu_1} - \epsilon_{\nu_2}}{K(\nu_1^{-2} - \nu_2^{-2})},
\ee
where  $K \approx 4.149 \times 10^9~\mathrm{\mu s~MHz^2~pc^{-1}~cm^{3}}$ is the dispersion constant \citep{handbook} and $\sigma_{\epsilon_{\nu}}$ is the rms timing uncertainty for frequency~$\nu$. 
(See the Appendix for more on this derivation and its effect on TOA perturbation.) 
The rms DM uncertainty, $\sigma_{\delta\DM}$, is then $\left<(\delta\DM)^2\right>^{1/2}$. Assuming that $\sigma_{\epsilon_{\nu}}$ is the same for the two halves of the LOFAR band and that $\sigma_{\delta\DM} = 3.7 \times 10^{-5}$~pc~cm$^{-3}$, we find that $\sigma_{\epsilon_{\nu}} \approx 2.3~\mu$s, \added{a factor of four smaller than the jitter uncertainty described above.} 

If the total TOA uncertainty is then the square root of the quadrature sum of the previous TOA uncertainty of 2.3~$\mu$s and the rms jitter of 10~$\mu$s, we can solve for the corrected rms DM uncertainty and find that $\sigma_{\delta\DM} = 1.7 \times 10^{-4}$~pc~cm$^{-3}$. This is the rms DM uncertainty determined over the whole frequency band. Next we calculate new DM uncertainties assuming that DM is derived from each half of the band separately (making sure to correct the TOA uncertainties for the change in the signal-to-noise ratio by a factor of $\sqrt{2}$ but no change in the jitter as it is roughly frequency-independent over a small frequency range; \citealt{FDJitter}), as is done by \DV to find the DM using measurements taken above and below 149~MHz. We split the full band into four and then used the centers of the bottom two frequency channels as our spot frequencies to determine the DM for data taken below 149~MHz. Correspondingly the centers of the top two frequency channels were used to determine the DM for data taken above 149~MHz. We found that $\sigma_{\delta\DM} = 3-5 \times 10^{-4}$~pc~cm$^{-3}$. This range of values is of similar order to those of $\sigma_{\DM(\nu)}$ discussed in \S\ref{sec:sf}. 
Our analysis describing the underestimation of the TOA uncertainties further strengthens the argument that differences in DM between the two frequency bands discussed in \DV are less significant than were shown. We do note, however, that while the amplitude of these uncertainties adds significantly to the two DM timeseries, this is a white-noise contribution in time and, therefore, cannot explain the systematic offsets between the two that are seen by \DV, \added{i.e., frequency-dependent DM causes non-white-noise shifts in the TOAs and so increasing the error bars does not remove the effect.}

We also estimated the contribution to the TOA uncertainty from scintillation noise, also known as the ``finite-scintle effect'' \citep{cwd+90}, one of three commonly-analyzed white-noise contributions to the TOA uncertainty on short timescales \citep{NG9WN,optimalfreq}. The scintillation timescale $\Dtd$ can be found when the structure function of the electromagnetic phase perturbation is equal to 1, or alternatively in terms of the DM structure function \citep{DMt},
\be 
D_{\DM}(\Dtd) = 1.47 \times 10^{-15}~\mathrm{(pc~cm^{-3})^2}~\left(\frac{\nu}{\mathrm{GHz}}\right)^2.
\ee
Given that they observe $D_\DM(\tau) = 3.1 \times 10^{-10}~ (\mathrm{pc~cm}^{-3})^2~(\tau/\mathrm{day})^{5/3}$, we found that $\Dtd = 5.7$~s. Due to the time-variability of the trailing components in the profile, we estimated the scattering timescale simply from NE2001 as $\taud = 0.4$~ms at 150~MHz, or an equivalent scintillation bandwidth of $\Dnud = 0.46$~kHz. The TOA uncertainty from scintillation noise is $\approx \taud/\sqrt{\niss}$, where $\niss$ is the number of scintles (``patches'' of increased intensity in the time-frequency plane), given by
\be 
\niss \approx \left(1 + \eta_t \frac{T}{\Dtd}\right)\left(1 + \eta_\nu \frac{B}{\Dnud}\right).
\ee
The filling factors $\eta_t, \eta_\nu$ are in the range 0.1-0.3 depending on the properties of the medium, and we have adopted a value of 0.2 for both \citep{cs10}. Given an observation time $T = 115$~minutes and a bandwidth $B = 71.5$~MHz, we found that contribution of scintillation noise to the TOA uncertainty is 0.4~$\mu$s, much smaller than either the template-fitting component or the jitter component, and therefore it should not factor into the error budget substantially.



\begin{figure*}[t!]
\centering
\includegraphics[width=1.0\textwidth]{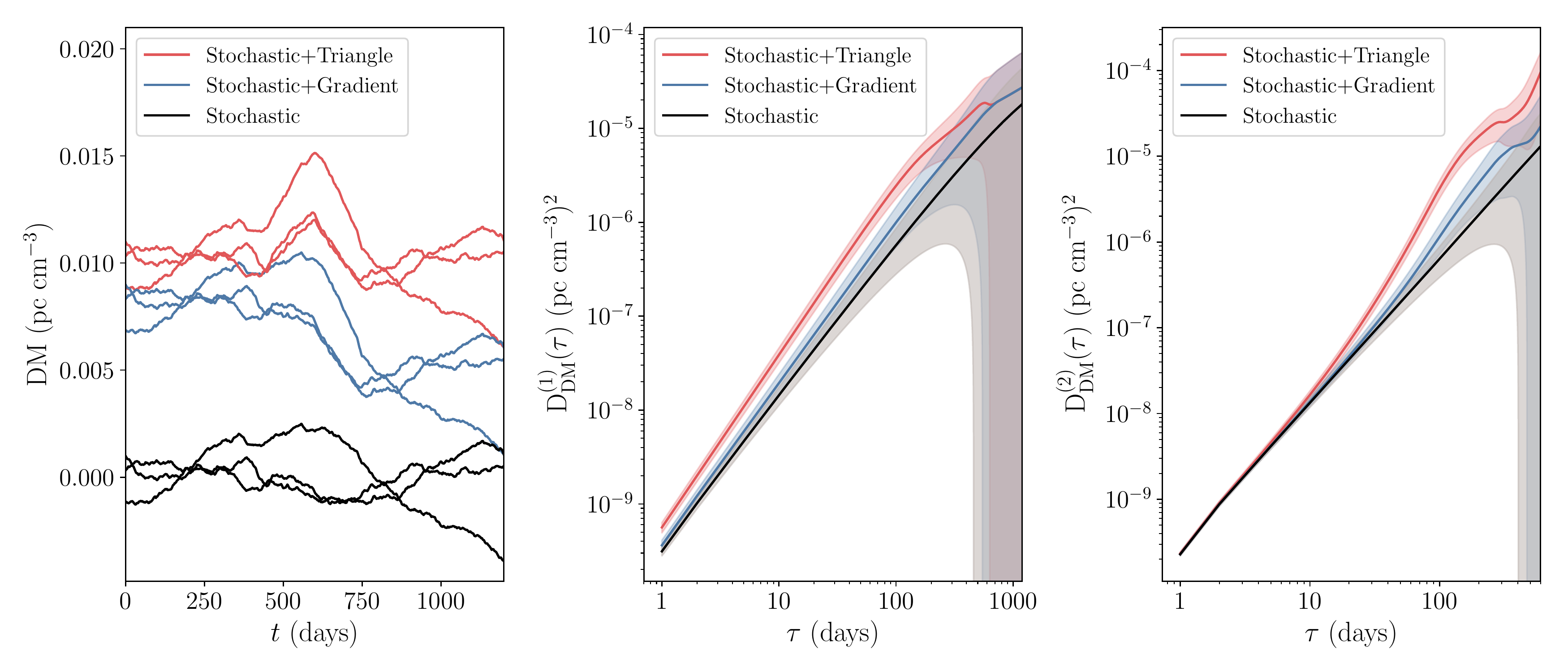}
  \caption{\footnotesize DM timeseries (left) along with their associated first- (middle) and second-order (right) structure functions. We plot the first three (arbitrary) timeseries (in order from bottom to top) we generated of the stochastic DM component (black) and added in a gradient (blue) or a triangle function representing an ESE (red). We offset the timeseries by $+0.005$~pc~cm$^{-3}$ between each set of three for visual clarity. The corresponding structure functions have the same colors, where the lines and the shaded regions represent the mean and standard deviation, respectively. See the text for more information.}
\label{fig:SFsim}
\end{figure*}

Lastly, we looked at the impact of polarization calibration errors on the data \citep{Stinebring+84}. \cite{Foster+15} have shown that polarization leakage can result in significant TOA uncertainties ($\sim$microseconds) for well-timed millisecond pulsars. \cite{NG11POL} showed that the stability of the Arecibo Observatory system varies quite dramatically between frequencies and epochs and therefore re-calibration must be performed per epoch. Given the errors, we performed an uncertainty analysis expanding upon the calibration procedure of \DV, who followed \cite{Noutsos+15}, where \PSR\ itself was used to test the calibration stability of the LOFAR antennas. \cite{Noutsos+15} stated that systematic uncertainties in the polarization leakage are of order 5--10\%.  

\DV performed observations at or close to transit, and therefore claim that imperfections in the calibration did not affect the analysis. However, \cite{Noutsos+15} showed that at hour angles far away from transit, the profile differences for \PSR\ can be as large as 30\% compared with at transit for the circularly polarized flux. While the observations \DV performed were closer to transit, we would still expect pulse shape deviations of order a few percent given the observation lengths alone.

Following \cite{Cordes+04}, the TOA uncertainty due to pulse shape changes from an incorrect absolute gain calibration is 
\be 
\sigma_{\rm pol} \sim 1~\mu\mathrm{s}~\left(\frac{\varepsilon}{0.1}\right)\left(\frac{\pi_{\mathrm{V}}}{0.1}\right)\left(\frac{W}{100~\mu\mathrm{s}}\right),
\label{eq:sigma_pol}
\ee
where we provide fiducial values as in \cite{NG9WN,optimalfreq} for the fractional gain error $\varepsilon$, degree of circular polarization $\pi_{\mathrm{V}}$, and pulse width $W$. Using Eq.~\ref{eq:sigma_pol} as a crude estimator, with a pulse with of $\sim 8$~ms, a circular polarization fraction of 9\% \citep{Noutsos+15}, and assuming $\varepsilon \sim 0.1$ (the gain error and polarization leakage are not entirely equivalent quantities but we take the fractional errors above as representative), then the component of the TOA uncertainty is $\sim 72~\mu$s, many times larger than the template-fitting errors. If polarization error yields a consistent offset/perturbation in the arrival times, then the net stochastic TOA uncertainty is zero and the frequency-dependent DM analysis should not be affected. However, given analyses such as that of \cite{NG11POL} regard system stability, we do not expect these calibration errors to be systematic alone. While \PSR\ is not a millisecond pulsar, we see that it is quite likely that polarization calibration errors on the order of microseconds or tens of microseconds are expected \citep[see again][]{Foster+15}, which since comparable to the template-fitting errors, should further be accounted for in dispersive-delay removal.

\subsection{Impact of a DM Gradient or Scattering Screen on the Structure Functions}

Any additional structure in the DM timeseries beyond that from a turbulent medium will increase the measured structure function \citep{DMt,NG9DM}.  For the following analysis, we will make a distinction between the first-order structure function $D_{\DM}^{(1)}(\tau)$ that we have implicitly discussed previously and the second-order structure function $D_{\DM}^{(2)}(\tau)$, both defined as follows:
\ba 
D_{\DM}^{(1)}(\tau) & \equiv & \left<\left[\DM(t+\tau) - \DM(t)\right]^2\right>, \nonumber \\
D_{\DM}^{(2)}(\tau) & \equiv & \left<\left[\DM(t+\tau) -2\DM(t) +\DM(t-\tau)\right]^2\right>.
\ea
While the first-order structure function will remove any constant term from the timeseries (e.g., the mean), the second-order structure function removes linear terms and can be used to detect discrete changes in any underlying linear trends in the DM timeseries \citep{DMt}. The latter can also be thought of as related to the curvature of the timeseries. For a Kolmogorov medium, it is proportional to the first-order structure function \citep[Appendix~A]{DMt}. Again, any additional structures beyond a turbulent medium seen in the timeseries will increase these measured quantities.

To understand the impact of a DM gradient or ESE on the structure functions of both orders, we performed simulations as in \cite{nonsimDM,DMt} of red-noise realizations of DM timeseries. We generated 10,000 realizations of 1200-day DM timeseries from a stochastic Kolmogorov medium for which the amplitude was set by the measured structure function $D_\DM^{(1)}(\tau) = 3.1 \times 10^{-10}~(\mathrm{pc~cm}^{-3})^2 (\tau/\mathrm{day})^{5/3}$. Next, we added a gradient with slope $2 \times 10^{-5}$~pc~cm$^{-3}$~day$^{-1}$ and of length 150 days into each timeseries. To avoid a discontinuity, we added a baseline value of $3 
\times 10^{-3}$~pc~cm$^{-3}$ to the higher side of the gradient (i.e., in total we added a slanted step function). Lastly, we added a triangle function representing an ESE into each stochastic realization (separate from the gradient), with slope $2 \times 10^{-3}$~pc~cm$^{-3}$~day$^{-1}$ and of length 150 days on either side.  Figure~\ref{fig:SFsim} (left panel) shows several of these timeseries.

As expected, we see in the first-order structure functions shown in the middle panel that there is an increase in the amplitude as well as a changing slope, though the mean stochastic + gradient structure function is within one standard deviation of the mean stochastic-only structure function. Note that because the length of the actual timeseries is only 1200 days, the range of $D_{\DM}^{(1)}$ at large lag varies 
significantly from the mean at lags greater than a few hundred days\footnote{Note that \DV calculate the standard deviation of the base-10 logarithm of the structure function\added{, whereas here we calculate the standard deviation of the structure function itself.}}. Since only a few increments (DM differences) contribute to the averages in the bins at large lags, we do expect a wide variation as we see in Figure~\ref{fig:SFsim} and in previous simulations of ours \citep{nonsimDM,DMt}.

If either a density gradient or an ESE is along the line of sight, then the stochastic Kolmogorov component of the measured first-order structure function should be lower in amplitude than previously reported, thereby lowering the measured SM or raising the scintillation timescale. While the slope of the mean first-order structure function for the stochastic + triangle simulations has a steeper slope, the ``realization errors'' on the structure function for the stochastic-only simulations can lead to both steeper or shallower slopes for a single measured structure function, which must be accounted for when performing these analyses to constrain the consistency of the spectrum with a Kolmogorov medium without bias \citep{DMt,NG9DM}.


Analyses of the second-order DM structure function have not been performed in the literature. We show the results of the analysis of our simulations in the right panel of Fig~\ref{fig:SFsim}. We notice two features. First, all three structure functions tend toward the same value at low time lags, which may then allow for a more robust estimate of the scintillation timescale (or alternatively the SM as in \citealt{Donner+19}) as per Appendix A of \cite{DMt}, and therefore the amplitude of the frequency-dependent DM. Second, we see that adding extra components to the DM timeseries on top of the purely stochastic term will produce a more pronounced increase in the value of the second-order structure function at a time lag of $\sim 150$~days, the timescale of the injected structures. There is still some slight overlap in the mean stochastic + gradient versus the mean stochastic-only second-order structure function.

\begin{figure}[t!]
\centering
\includegraphics[width=0.5\textwidth]{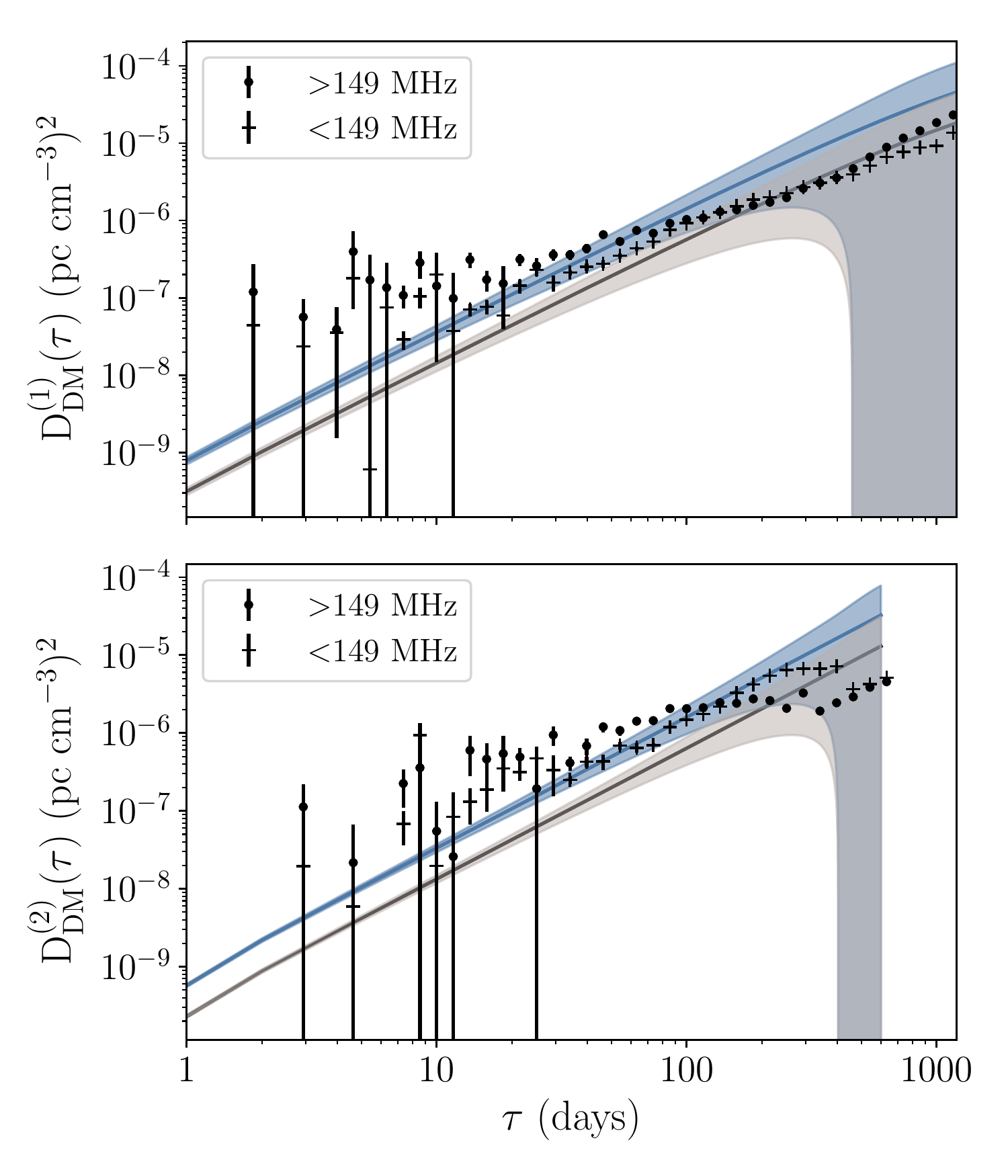}
  \caption{\footnotesize First- (top) and second-order (bottom) structure functions for the two DM timeseries estimated for different frequencies (the symbols representing the two frequency ranges) as shown in \DV. The gray region denotes the simulations as shown in Figure~\ref{fig:SFsim} for the stochastic-only medium with the amplitude set by that estimated in \DV. The blue region denotes those scaled by a factor of 2.5, showing greater consistency with both structure functions.
  }
\label{fig:sfdataplot}
\end{figure}

To test our simulations, we used the values of the two DM timeseries estimated for different frequencies as provided in \DV
to calculate both the first- and second-order structure functions for each, shown in Figure~\ref{fig:sfdataplot}. 
The gray regions are the same as in Figure~\ref{fig:SFsim} for the stochastic-only simulations, again scaled to the amplitude they estimate of $D_\DM^{(1)}(\tau) = 3.1 \times 10^{-10}~(\mathrm{pc~cm}^{-3})^2 (\tau/\mathrm{day})^{5/3}$. Recall that no white noise has been added to our simulations, and therefore there is no flattening of the simulated structure functions as compared with the values estimated from the data. 
We see significant spread in the values of both structure functions at low lags, due partially to the white noise and partially from the irregular sampling at small lags. 

For the first-order structure function, we see no increase in the values for $\tau \gtrsim 100$~days indicative of a discrete structure or gradient as analyzed in our simulations (see again Figure~\ref{fig:SFsim}). Additionally, we see that the gray regions for the second-order structure function do not well-represent the data for most lags, and we also do not see the equivalent increase at large lags if a discrete structure was present. 

By increasing the overall amplitude of the simulated structure functions, we see greater agreement with the data, which can be used to extract information about the medium. As fitting of the structure functions, or even the increments, can be complex as noted previously \citep{DMt,NG9DM}, we decided instead to visually increase the values only for demonstrative purposes about the utility of these estimates. We increased the simulated structure functions by a factor of 2.5, which are seen in the blue shaded regions in Figure~\ref{fig:sfdataplot}. The greater consistency with the data points is easily visible. Since $\sigma_{\DM(\nu)} \propto \SM \propto C_n^2 \propto D_{\DM}^{(1)}$ (see Eq.~\ref{eq:sigmaDMnu_uniform}, and also \citealt{css16}, \citealt{DMt}), then a factor of 2.5 increase in the structure function translates to the same increase in $\sigma_{\DM(\nu)}$. 

However, the observed rms difference we calculated between the two DM timeseries was $0.7 \times 10^{-3}$~pc~cm$^{-3}$ (see also Figure~6 of \citealt{Donner+19}), which was in agreement with the prediction for a uniform Kolmogorov medium discussed in \S\ref{sec:sf}. Since the rms difference\footnote{Note that while typically the structure function is proportional to the variance of a time series at a given lag, in this case we are looking at the proportionality with the rms frequency-dependent DM rather than the rms of the DM timeseries.} in the frequency-dependent DM is increasing by a factor of 2.5 due to an increase in the estimated $D_{\DM}^{(1)}$, or alternatively the SM, then we find disagreement with the initial assumption of the line of sight being a uniform Kolmogorov medium. 

One common model for describing the ISM along the line of sight is not as uniform but with a thin scattering screen model since the mathematics is simpler and the resulting quantities in a calculation depend only on geometric terms \citep{cpl86,cr98}; the geometry of the screen is independent from the spectral index (i.e., Kolmogorov or otherwise). Such a model can also physically describe regions of larger-scale overdense structure in the ISM. Rather than assuming that the medium along the line of sight is entirely uniform, we can quantify what effect a factor of 2.5 increase in the structure function implies if we assume the thin-screen model. In Eq.~\ref{eq:sigmaDMnu_uniform}, we provided the geometric factor $G_{11/3} = 0.145$ for a uniform medium via \citet{css16}. If instead we assume a screen at distance $D_s$ from the pulsar, at distance $D$, then the geometric factor becomes $G_{11/3} = \left[x(1-x)\right]^{5/6}$, where $x = D_s/D$ is the screen's fractional distance between the pulsar and us. Note that the function is symmetric about the halfway ($x=0.5$) point. Assuming a screen describes the line of sight allows us to simultaneously increase the SM while decreasing the value of $G_{11/3}$ to keep $\sigma_{\DM(\nu)}$ fixed at its observed value. Taking the observed DM difference rms and using Eq.~\ref{eq:sigmaDMnu_uniform} to instead determine $G_{11/3}$ and therefore $x$, we found that $x = 0.045$ or $x = 0.955$, i.e., the screen is at a distance 0.1~kpc from either the pulsar or the Earth. In the case of the latter, this nominal distance lies a factor of a few beyond that of the Local Bubble \citep{frs11}. An $x \approx 0.98$ would place the screen in the rough appropriate location ($\sim 40$~pc), which in turn decreases $G_{11/3}$ or increases the structure functions by an additional factor of 2, which starts to come in tension with the values shown in Figure~\ref{fig:sfdataplot}. However, since the pulsar's distance is uncertain due to estimation with the NE2001 model, then the Local Bubble could still be a plausible location of a screen.

Since $G_{11/3}$ and therefore $\sigma_{\DM(\nu)}$ increases as $x$ tends toward the midpoint of the line of sight, we can rule out any screen structure much closer than the endpoints of the line of sight since the estimated structure functions would be in greater tension with those observed. Our method demonstrates that we can simultaneously use the variations in the DM timeseries taken at different frequencies along with the variations of the differences between the two DM timeseries to constrain the location of an assumed screen quite strongly. Future observations of this and other pulsars taken at widely separated frequencies with high cadence may be able to jointly model even higher-order structure functions for even further constraining power.

Due to irregular sampling in DM timeseries, common for many pulsar observations, it may be preferred to analyze the individual second-order increments ($\DM(t+\tau) -2\DM(t) +\DM(t-\tau)$, which are squared and averaged over to obtain $D_{\DM}^{(2)}$) as was performed for PSR~B1534+12 in \cite{DMt} than the second-order structure function as performed here. However, the high-cadence of observations in \DV combined with the duration has allowed us to directly calculate the second-order structure function rather than simply the increments. Continued observations, especially with a dense high-cadence program, could help constrain the values of both structure functions at smaller lags.


 Note that the large transverse velocity means that the motion of the pulsar across the sky is fairly straight (e.g., the parallax motion is small, see trajectory plots in \citealt{NG9DM}). The quasi-periodic variations in the DM timeseries that deviate from a purely power-law spectrum, seen in \cite{NG11SW} due to the line-of-sight crossing correlated spatial DM fluctuations, should therefore be small and not impact the power spectrum/structure function significantly. Put another way, while the assumption of the line of sight crossing independent DM fluctuations is broken and we expect short-term rapid variations in the DM that might impact the measured structure function, we expect this change to be negligible for this pulsar.

In conclusion, the consistency of the structure functions 
with a Kolmogorov model over the range of time and spatial scales observed suggests that the medium is turbulent down to 10s of days or $\sim$1~AU (for an assumed lens at distance 1.1~kpc). As discussed, below this scale the structure function appears dominated by white-noise fluctuations given the cadence of observations, which causes the observed structure function/spectrum to become shallower. The $L=20$~AU lens size (equivalent to the 300~day timescale) should be seen as a steepening in the structure function as shown in Figure~\ref{fig:SFsim}. In addition, any truly stochastic process with a spectrum steeper than that of a Kolmogorov medium, which could then produce strong refractive effects \citep{cpl86}, is also unseen in the structure functions in Figure~\ref{fig:sfdataplot}.




\section{Impact on Precision Timing}
\label{sec:timing}

Temporal and frequency-dependent DM variations will have significant impacts on high-precision pulsar timing experiments, including the efforts to detect low-frequency gravitational waves. While these topics have been discussed in the literature (see e.g., \citealt{You+07a,Keith+2013,Lee+14,DMt,css16}), we will take the analysis of \DV further and discuss the impact in the context of our \added{accounting of the various noise processes in the real data.}




\DV\ estimated the limiting TOA precision by considering observations at~1.4~GHz with a relatively limited 250~MHz bandwidth (20\% bandwidth).  However, current pulsar backend systems can process up to 800~MHz of bandwidth \citep[see e.g., the Appendix of][]{global1713} and techniques have been developed to process even larger bandwidths \citep[e.g.,][]{UWB} while simultaneously compensating for the DM \citep{Pennucci+14}.  More generally, a common practice is to conduct simultaneous or near-simultaneous observations at two frequencies (e.g., the North American Nanohertz Observatory for Gravitational Waves, NANOGrav, uses combinations of 0.8~GHz and 1.4~GHz for some pulsars and 1.4~GHz to 2.3~GHz for some others; \citealt{NG11}), from which even higher precision DM estimates can be obtained.

When only purely dispersive delays factor into TOAs measured at two spot frequencies, the rms TOA is given by (see the Appendix but also \citealt{Lee+14}, Eqn.~12; \citealt{optimalfreq}, Table~1),
\be 
\sigma_{\delta t_\infty} = \left(\frac{\sigma_{\epsilon_{\nu_1}}^2 + r^4 \sigma_{\epsilon_{\nu_2}}^2}{(r^2-1)^2}\right)^{1/2},
\label{eq:rmsTOA}
\ee
where again $\sigma_{\epsilon_{\nu}}$ is the rms timing uncertainty for frequency~$\nu$. This timing uncertainty includes all sources of white noise that affect the TOA estimation, such as from radiometer noise or jitter \citep[see e.g.,][]{optimalfreq}.  A critial aspect of Eq.~\ref{eq:rmsTOA} is that the rms TOA is  dependent only on $r$ and not a specific frequency.  It does not matter whether the timing measurements are acquired at low or high frequencies so long as the individual frequency-channel TOA uncertainties are the same for a given value of~$r$.

As an illustration, taking $r \sim 2$ and setting $\sigma_{\epsilon_{\nu}} = \sigma_{\epsilon_{\nu_1}} = \sigma_{\epsilon_{\nu_2}}$, then $\sigma_{\delta t_\infty} = 1.4 \sigma_{\epsilon_{\nu}}$. Therefore, in cases where the median TOA uncertainty from finite pulse signal-to-noise ratio is small (i.e., well below 1~$\mu$s), such as reported per-pulsar and per-band in the NANOGrav 11-Year Data Set \citep{NG11}, then the requirement for sub-microsecond precision timing is met. Several other sources of error, such as from pulse phase jitter, are known to be much smaller than this limit \citep{FDJitter}. Therefore, using low-frequency timing data to increase $r$ will quantitatively improve the timing of many pulsars as long as the TOA uncertainties for pulses at those frequencies are low enough and unmitigated ISM effects are small, e.g., typically for pulsars with lower DM values \citep{optimalfreq}.

As an extreme case, we considered $\sigma_{\epsilon_{\nu_2}} \to 0$.  In this case, the overall timing precision will be dominated by the precision at the lower frequency, $\sigma_{\delta t_\infty} \approx \sigma_{\epsilon_{\nu_1}}/(r^2-1)$.  For a sufficiently large value of~$r$, the timing precision due to DM uncertainty or variations could be made negligible.  As a specific example, with relevance to the data presented by \DV, we considered $r \approx 10$, equivalent to $\nu_1 \approx 140$~MHz and $\nu_2 \approx 1400$~MHz.  In this hypothetical example, a timing precision $\sigma_{\delta t_\infty} \approx 10$~ns could be obtained, if $\sigma_{\epsilon_{\nu_1}} \approx 1\,\mu\mathrm{s}$.  We chose this illustration because a timing precision of order 10~ns is comparable to the expected precision required for the study of gravitational waves. 

This timing improvement of course neglects the uncertainties due to frequency-dependent DM as discussed by \citet{css16}. The amplitude of the differences in DM between frequencies also depends on $r$ (as well as the specific frequency choices) but the net effect on the overall timing precision can be quantified and then built into noise models, e.g., via covariance matrices such as constructed in \cite{optimalfreq}.  \citet[][see Figures~7 and~8]{css16} show that for pulsars with $\DM \lesssim 30$~pc~cm$^{-3}$, combining data from the 100~MHz and 2~GHz regimes will yield TOA errors that still meet the requirement of sub-microsecond precision, especially as current and future telescopes come online, drastically reducing the template-fitting errors from finite pulse signal-to-noise ratio. Many pulsars used in precision-timing experiments have DM values in this range \citep{IPTADR1}. \citet{css16} also discussed fitting a wide range of frequencies versus only two spot values; as expected, the increase in frequency coverage can in many cases improve the timing precision over the case where only two spot frequencies are used, which is no longer the case for many modern precision-pulsar-timing experiments \citep{IPTADR1}.

Unmitigated chromatic delays will also add to the TOA uncertainties \citep[Table~1]{optimalfreq} and cause frequency-dependent excess noise in the timing residuals \citep{IPTADR1noise,NG9EN}. However, even for the pulsar with the highest \hbox{DM} millisecond pulsar used in precision timing, PSR~J1903+0327 ($\DM \approx 300$~pc~cm$^{-3}$), observed between roughly 1.1 and 2.5~GHz, the long-term rms residual is $4~\mu$s, though a significant portion of that rms is again from frequency-dependent excess noise \citep{NG9EN}, which future timing methodologies might be able to mitigate partially \citep{sc17}.

Lastly, we expand on the discussion in \DV that at higher frequencies, longer-term DM variations are of particular importance to take into account for timing data. Given that the frequency-dependent DM comes from differences in ray-path averaging whereas the trends in DM come from the relative Earth-pulsar motion along the line of sight, we expect the longer-term DM variations to track each other between frequencies, regardless of frequency; \DV also agree with this given their analysis. However, in terms of the overall timing, it is the lower frequencies that are impacted much more heavily since the dispersive delays are weighted by $\nu^{-2}$. At frequencies much higher than typically used in precision timing, the dispersive delay becomes small and thus any changes in the dispersive delay are also small; as an extreme, X-ray pulsar data do not require DM corrections of any kind.

As an additional consideration, it is important to remember that short-term DM variations that are improperly corrected for can contribute heavily to the overall rms timing, and may not contribute as simple white noise \citep[e.g.,][]{nonsimDM}. Many pulsars in precision-timing experiments show very rapid timescales for DM to vary \citep{NG9DM}, including from the solar wind (\citealt{You+07b}; \citealt{NG11SW}; see also \citealt{Howard+16} for the study of a coronal mass ejection with a slow-period pulsar) or structures in the ISM \citep[e.g.,][]{Fonseca+14,Coles+15,secondISMevent}.


\section{Discussion}
\label{sec:discussion}

In this work, we have described the requirements necessary for observational tests separating the effects of frequency-dependent DM from refractive lensing. The observations shown in \DV provide an excellent test of frequency-dependent DM as laid out by \citet{css16}. While the theoretical treatment considers the ability to perfectly measure DM at a given spot frequency,  going forward, analyses of observations of the kind reported in \DV must account for the wide range of different TOA uncertainty components and ISM propagation effects presented in this work. We showed here that even for a canonical pulsar versus a millisecond pulsar, the recent work done in precision-timing experiments has become relevant given the timing quality of the pulsars and the instrumentation used to access to new types of observations, e.g., low-frequency observations via LOFAR as described here. And while canonical pulsars may not provide constraints on the same tests of fundamental physics as millisecond pulsars, their use in studying variations in the ionized ISM along many different lines of sight \citep[e.g.,][]{Petroff+13} will be unparalleled given the greater population of them over millisecond pulsars, especially if observations covering a large frequency ratio can be leveraged. 

While we have focused our analyses on the plasma lens as described by \DV, we have not discussed the potential causes of the light echoes as seen by \MH. They described the similarities between their pulse profiles and those seen in PSR B0531+21 (the Crab Pulsar), though those profile variations are attributed to structure in the local environment in the nebula surrounding the pulsar \citep{bwv00,lpg01}. While an interstellar lens would produce a negligible effect if very close to the pulsar, as shown in Figure~\ref{fig:lens_perturbations}, the model proposed by \citet{bwv00} involves the pulsar traversing near material (a prism geometry) close enough that emission at different frequencies passes through different electron content at different times (a true dispersive delay, whereas they argue that the refractive geometric delay will be significantly smaller) while that of \citet{lpg01} involves reflections of the images. Note that the fact that the Crab Pulsar has both a main pulse and interpulse means that propagation variations will affect both in the same way, making it easier to disentangle from intrinsic profile shape variations. Both of the proposed mechanisms could be examined in more detail with respect to the pulse profiles shown in \MH. If in a combined analysis the variability is also tied to the DM variations shown in \DV, then the total data set will provide an excellent probe of the material local to \PSR, though such variations will then need to be disentangled from the variability expected from a turbulent medium with which we have shown consistency.

Identifying ESEs or other ``ISM events'' seen in pulsar timing data in near-real time will allow for more intensive follow-up observations, including a higher cadence of observations over many frequencies and using different observatories worldwide, especially if dynamic spectra with resolved scintles can be obtained \citep{Hewish80,Stinebring+01}. The characteristic timescale and bandwidth can be used to constrain the location of a lensing structure \citep{cr98} while the drift rate (``rotation'') of the scintles provides the component of the refractive angle in the direction of the pulsar's motion \citep{hwg85,cpl86}, providing partial information on the geometric time delay. Interferometric observations can help constrain the changing position and sizes of the pulsar image \citep{bn85}, or possibly multiple images \citep{cw86,Cordes+17}, providing additional constraints on the lensing geometry over the line of sight. Any such additional observations will allow us to resolve small-scale structure in the ISM and probe the Galactic population of these lenses.

\begin{acknowledgements}
The NANOGrav Project receives support from NSF Physics Frontiers Center award number 1430284. Part of this research was carried out at the Jet Propulsion Laboratory, California Institute of Technology, under a contract with the National Aeronautics and Space Administration. Part of this research has made use of the database of published pulse profiles maintained by the European Pulsar Network, available at: \url{http://www.jb.man.ac.uk/research/pulsar/Resources/epn/}.

\end{acknowledgements}

\begin{appendix}

\section{DM Estimation with Additional Chromatic Errors}
\label{sec:appendix}

As a useful reference, here we will describe DM estimation from observations taken at two spot frequencies with additional chromatic errors following the formalism of \cite{nonsimDM} and \citet{css16}. We can write the TOA at a particular frequency $\nu$ as the infinite-frequency arrival time plus the dispersive delay term. For this calculation, we will also include measurement errors $\epsilon_\nu$ and a chromatic (frequency-dependent) timing perturbation $t_{C,\nu}$, such that
\be
t_\nu = t_\infty + \frac{K\DM}{\nu^2} + t_{C,\nu} + \epsilon_\nu.
\ee
Here, $K \approx 4.149 \times 10^9~\mathrm{\mu s~MHz^2~pc^{-1}~cm^{3}}$ is the dispersion constant in observationally convenient units \citep{handbook}. We estimate the DM by taking TOAs at two frequencies $\nu_1$ and $\nu_2$ and calculating
\be
\DMhat = \frac{t_{\nu_1} - t_{\nu_2}}{K(\nu_1^{-2} - \nu_2^{-2})}.
\ee
 As in the main text, we will define the frequency ratio $r \equiv \nu_2/\nu_1$ with $\nu_1<\nu_2$. The estimated infinite-frequency arrival time can then be written in one of two ways as
\begin{alignat}{2}
\hat{t}_\infty & = t_{\nu_1} - \frac{K\DMhat}{\nu_1^2} &{}={}& t_\infty + \frac{K(\DM-\DMhat)}{\nu_1^2} + t_{C,\nu_1}  + \epsilon_{\nu_1} \nonumber \\
& = t_{\nu_2} - \frac{K\DMhat}{\nu_2^2} &{}={}& t_\infty + \frac{K(\DM-\DMhat)}{\nu_2^2} + t_{C,\nu_2}  + \epsilon_{\nu_2} 
\end{alignat}

We will now solve for the DM difference. Substituting the measured TOAs $t_\nu$ into the equation for $\DMhat$ and subtracting from the true DM, we have
\ba
\delta\DM \equiv \DM - \DMhat & = & \DM -\frac{t_{\nu_1} - t_{\nu_2}}{K(\nu_1^{-2} - \nu_2^{-2})} \nonumber \\
& = & - \frac{t_{C,\nu_1} - t_{C,\nu_2} + \epsilon_{\nu_1} - \epsilon_{\nu_2}}{K(\nu_1^{-2} - \nu_2^{-2})}.
\label{eq:deltaDM}
\ea
The TOA perturbation will be
\ba
\delta t_\infty \equiv t_\infty - \hat{t}_\infty & = & - \frac{K(\DM-\DMhat)}{\nu_2^2} - t_{C,\nu_2} - \epsilon_{\nu_2} \nonumber \\
& = & \frac{- r^2 t_{C,\nu_2}  - r^2 \epsilon_{\nu_2} + t_{C,\nu_1} + \epsilon_{\nu_1} }{r^2 - 1}.
\ea
When the chromatic offsets are zero, we arrive at simply
\be
\delta t_\infty = \frac{\epsilon_{\nu_1} - r^2 \epsilon_{\nu_2} }{r^2 - 1},
\label{eq:dt}
\ee
which agrees with Eq.~21 in \citet{css16} assuming the frequency-dependent DM term is zero.

Eq.~\ref{eq:dt} provides the timing offset but one must consider the TOA uncertainty, $\sigma_{\delta t_\infty}$, from the variance
\be 
\sigma_{\delta t_\infty}^2 = \left<\delta t_\infty^2\right> = \left<\left(\frac{\epsilon_{\nu_1} - r^2 \epsilon_{\nu_2}}{r^2 - 1}\right)^2\right>.
\label{eq:variance}
\ee
We have assumed here that $\left<\delta t_\infty\right> = 0$, which will be true if the errors $\epsilon_\nu$ are Gaussian distributed. If they are, and with variance $\sigma_{\epsilon_\nu}^2$, then the sum of the two terms in the numerator of Eq.~\ref{eq:variance} will be Gaussian distributed, which when squared will then be chi-squared distributed. Taking the expected value of the resultant quantity yields
\be 
\sigma_{\delta t_\infty}^2 = \frac{\sigma_{\epsilon_{\nu_1}}^2 + r^4 \sigma_{\epsilon_{\nu_2}}^2}{(r^2-1)^2}.
\ee
If the TOA uncertainties are the same at both frequencies such that $\sigma_{\epsilon_{\nu_1}} = \sigma_{\epsilon_{\nu_2}} = \sigma_{\epsilon_\nu}$, then we have
\be 
\sigma_{\delta t_\infty} = \sqrt{\left<\delta t_\infty^2\right>} = \sigma_{\epsilon_\nu} \left(\frac{r^4 +1}{r^4 - 2r^2 +1}\right)^{1/2}.
\ee

\end{appendix}



\begin{thebibliography}{}

\bibitem[Armstrong et al.(1995)]{Armstrong+95} Armstrong, J.~W., Rickett, B.~J., \& Spangler, S.~R.\ 1995, \apj, 443, 209 


\bibitem[Arzoumanian et al.(2018)]{NG11} Arzoumanian, Z., Brazier, A., Burke-Spolaor, S., et al.\ 2018, \apjs, 235, 37 

\bibitem[Backer et al.(1993)]{Backer+93} Backer, D.~C., Hama, S., van Hook, S., \& Foster, R.~S.\ 1993, \apj, 404, 636 

\bibitem[Backer et al.(2000)]{bwv00} Backer, D.~C., Wong, T., \& Valanju, J.\ 2000, \apj, 543, 740 

\bibitem[Bilous et al.(2016)]{Bilous+16} Bilous, A.~V., Kondratiev, V.~I., Kramer, M., et al.\ 2016, \aap, 591, A134 

\bibitem[Blandford \& Narayan(1985)]{bn85} Blandford, R., \& Narayan, R.\ 1985, \mnras, 213, 591 

\bibitem[Clegg et al.(1998)]{Clegg+98} Clegg, A.~W., Fey, A.~L., \& Lazio, T.~J.~W.\ 1998, \apj, 496, 253 

\bibitem[Cognard et al.(1993)]{Cognard+93} Cognard, I., Bourgois, G., Lestrade, J.-F., et al.\ 1993, \nat, 366, 320 

\bibitem[Coles et al.(2015)]{Coles+15} Coles, W.~A., Kerr, M., Shannon, R.~M., et al.\ 2015, \apj, 808, 113

\bibitem[Cordes \& Downs(1985)]{cd85}
	Cordes, J. M., \& Downs, G.~S.  1985, 
    \apjs, 59, 343

    
\bibitem[Cordes \& Wolszczan(1986)]{cw86}
    Cordes, J. M., \& Wolszczan, A.  1986,
    \apj, 307, L27
    
\bibitem[Cordes et al.(1986)]{cpl86} Cordes, J.~M., Pidwerbetsky, A., \& Lovelace, R.~V.~E.\ 1986, \apj, 310, 737 
    
\bibitem[Cordes et al.(1990)]{cwd+90} Cordes, J.~M., Wolszczan, A., Dewey, R.~J., Blaskiewicz, M., \& Stinebring, D.~R.\ 1990, \apj, 349, 245 
    
\bibitem[Cordes \& Rickett(1998)]{cr98} Cordes, J. M. \& Rickett, B. J.\ 1998, \apj, 507, 846

\bibitem[Cordes \& Lazio(2002)]{NE2001} Cordes, J.~M., \& Lazio, T.~J.~W.\ 2002, arXiv:astro-ph/0207156

\bibitem[Cordes et al.(2004)]{Cordes+04} Cordes, J.~M., Kramer, M., Lazio, T.~J.~W., et al.\ 2004, \nar, 48, 1413 

\bibitem[Cordes \& Shannon(2010)]{cs10} Cordes, J.~M., \& Shannon, R.~M.\ 2010, arXiv:1010.3785 

\bibitem[Cordes et al.(2016)]{css16}
    Cordes, J.~M., Shannon, R.~M., \& Stinebring, D.~R.\ 2016, \apj, 817, 16 
    
\bibitem[Cordes et al.(2017)]{Cordes+17} Cordes, J.~M., Wasserman, I., Hessels, J.~W.~T., et al.\ 2017, \apj, 842, 35 
    
\bibitem[Craft(1970)]{CraftThesis} Craft, H.~D., Jr.\ 1970, Ph.D.~Thesis, Cornell University

\bibitem[Demorest(2007)]{DemorestThesis} Demorest, P.~B.\ 2007, Ph.D.~Thesis,  University of California, Berkeley

\bibitem[Demorest et al.(2013)]{NG5} Demorest, P.~B., Ferdman, R.~D., Gonzalez, M.~E., et al.\ 2013, \apj, 762, 94 

\bibitem[Desvignes et al.(2016)]{Desvignes+16} Desvignes, G., Caballero, R.~N., Lentati, L., et al.\ 2016, \mnras, 458, 3341 


\bibitem[Dolch et al.(2014)]{global1713} Dolch, T., Lam, M.~T., Cordes, J., et al.\ 2014, \apj, 794, 21 

\bibitem[Donner et al.(2019)]{Donner+19} Donner, J.~Y., Verbiest, J.~P.~W., Tiburzi, C., et al.\ 2019, \aap, 624, A22


\bibitem[Dunning et al.(2015)]{UWB} Dunning, A., Bowen, M., Bourne, M., Hayman, D., \& Smith, S.~L.\ 2015. {\it An ultra-wideband dielectrically loaded quad-ridged feed horn for radio astronomy}. In 2015 IEEE-APS Topical Conference on Antennas and Propagation in Wireless Communications (APWC) (pp. 787-790).

\bibitem[Fiedler et al.(1987)]{Fiedler+87} Fiedler, R.~L., Dennison, B., Johnston, K.~J., \& Hewish, A.\ 1987, \nat, 326, 675 

\bibitem[Fiedler et al.(1994)]{Fiedler+94} Fiedler, R., Pauls, T., Johnston, K.~J., \& Dennison, B.\ 1994, \apj, 430, 595 

\bibitem[Fonseca et al.(2014)]{Fonseca+14} Fonseca, E., Stairs, I.~H., \& Thorsett, S.~E.\ 2014, \apj, 787, 82 

\bibitem[Foster \& Cordes(1990)]{fc90} Foster, R.~S., \& Cordes, J.~M.\ 1990, \apj, 364, 123 

\bibitem[Foster et al.(2015)]{Foster+15} Foster, G., Karastergiou, A., Paulin, R., et al.\ 2015, \mnras, 453, 1489 

\bibitem[Frisch et al.(2011)]{frs11} Frisch, P.~C., Redfield, S., \& Slavin, J.~D.\ 2011, \araa, 49, 237 


\bibitem[Gentile et al.(2018)]{NG11POL} Gentile, P.~A., McLaughlin, M.~A., Demorest, P.~B., et al.\ 2018, \apj, 862, 47 

\bibitem[Hassall et al.(2012)]{Hassall+12} Hassall, T.~E., Stappers, B.~W., Hessels, J.~W.~T., et al.\ 2012, \aap, 543, A66 
\bibitem[Hewish(1980)]{Hewish80} Hewish, A.\ 1980, \mnras, 192, 799 

\bibitem[Hewish et al.(1985)]{hwg85}
	Hewish, A., Wolszczan, A., \& Graham, D.~A.  1985, 
    \mnras, 213, 167

\bibitem[Hobbs et al.(2004)]{Hobbs+2004} Hobbs, G., Lyne, A.~G., Kramer, M., Martin, C.~E., \& Jordan, C.\ 2004, \mnras, 353, 1311 

\bibitem[Howard et al.(2016)]{Howard+16} Howard, T.~A., Stovall, K., Dowell, J., Taylor, G.~B., \& White, S.~M.\ 2016, \apj, 831, 208 

\bibitem[Jones et al.(2017)]{NG9DM}
    Jones, M.~L., McLaughlin, M.~A., Lam, M.~T., et al.\ 2017, \apj, 841, 125 

\bibitem[Keith et al.(2013)]{Keith+2013} Keith, M.~J., Coles, W., Shannon, R.~M., et al.\ 2013, \mnras, 429, 2161 

\bibitem[Lam et al.(2015)]{nonsimDM} Lam, M.~T., Cordes, J.~M., Chatterjee, S., \& Dolch, T.\ 2015, \apj, 801, 130 


\bibitem[Lam et al.(2016a)]{NG9WN} Lam, M.~T., Cordes, J.~M., Chatterjee, S., et al.\ 2016a, \apj, 819, 155 

\bibitem[Lam et al.(2016b)]{DMt} Lam, M.~T., Cordes, J.~M., Chatterjee, S., et al.\ 2016b, \apj, 821, 66 

\bibitem[Lam et al.(2017)]{NG9EN} Lam, M.~T., Cordes, J.~M., Chatterjee, S., et al.\ 2017, \apj, 834, 35 

\bibitem[Lam(2017)]{PyPulse} Lam, M.~T.\ 2017, Astrophysics Source Code Library, ascl:1706.011 

\bibitem[Lam et al.(2018a)]{optimalfreq} Lam, M.~T., McLaughlin, M.~A., Cordes, J.~M., Chatterjee, S., \& Lazio, T.~J.~W.\ 2018a, \apj, 861, 12 

\bibitem[Lam et al.(2018b)]{secondISMevent} Lam, M.~T., Ellis, J.~A., Grillo, G., et al.\ 2018b, \apj, 861, 132 

\bibitem[Lam et al.(2019)]{FDJitter} Lam, M.~T., McLaughlin, M.~A., Arzoumanian, Z., et al.\ 2019, \apj, 872, 193

\bibitem[Lee et al.(2014)]{Lee+14} Lee, K.~J., Bassa, C.~G., Janssen, G.~H., et al.\ 2014, \mnras, 441, 2831 

\bibitem[Lentati et al.(2016)]{IPTADR1noise} Lentati, L., Shannon, R.~M., Coles, W.~A., et al.\ 2016, \mnras, 458, 2161 

\bibitem[Levin et al.(2016)]{Levin+16} Levin, L., McLaughlin, M.~A., Jones, G., et al.\ 2016, \apj, 818, 166 

\bibitem[Lorimer \& Kramer(2012)]{handbook} Lorimer, D.~R., \& Kramer, M.\ 2012, Handbook of Pulsar Astronomy, by D.~R.~Lorimer , M.~Kramer, Cambridge, UK: Cambridge University Press, 2012

\bibitem[Lyne et al.(2001)]{lpg01} Lyne, A.~G., Pritchard, R.~S., \& Graham-Smith, F.\ 2001, \mnras, 321, 67 

\bibitem[Madison et al.(2018)]{NG11SW} Madison, D.~R., Cordes, J.~M., Arzoumanian, Z., et al.\ 2019, \apj, 872, 150

\bibitem[Maitia et al.(2003)]{mlc03} Maitia, V., Lestrade, J.-F., \& Cognard, I.\ 2003, \apj, 582, 972 

\bibitem[Manchester et al.(2013)]{Manchester+13} Manchester, R.~N., Hobbs, G., Bailes, M., et al.\ 2013, \pasa, 30, e017 

\bibitem[Michilli et al.(2018)]{Michilli+18} Michilli, D., Hessels, J.~W.~T., Donner, J.~Y., et al.\ 2018, \mnras, 476, 2704 

\bibitem[Noutsos et al.(2015)]{Noutsos+15} Noutsos, A., Sobey, C., Kondratiev, V.~I., et al.\ 2015, \aap, 576, A62 

\bibitem[Os{\l}owski et al.(2011)]{Oslowski+11} Os{\l}owski, S., van Straten, W., Hobbs, G.~B., Bailes, M., \& Demorest, P.\ 2011, \mnras, 418, 1258 

\bibitem[Pennucci et al.(2014)]{Pennucci+14} Pennucci, T.~T., Demorest, P.~B., \& Ransom, S.~M.\ 2014, \apj, 790, 93 

\bibitem[Pennucci(2015)]{PennucciThesis} Pennucci, T.~T.\ 2015, Ph.D.~Thesis, University of Virginia

\bibitem[Petroff et al.(2013)]{Petroff+13} Petroff, E., Keith, M.~J., Johnston, S., van Straten, W., \& Shannon, R.~M.\ 2013, \mnras, 435, 1610 

\bibitem[Phillips \& Wolszczan(1991)]{pw91} Phillips, J.~A., \& Wolszczan, A.\ 1991, \apjl, 382, L27 

\bibitem[Phillips \& Wolszczan(1992)]{pw92} Phillips, J.~A., \& Wolszczan, A.\ 1992, \apj, 385, 273 

\bibitem[Rankin \& Roberts(1971)]{rr71} Rankin, J.~M., \& Roberts, J.~A.\ 1971, The Crab Nebula, 46, 114 

\bibitem[Ramachandran et al.(2006)]{Ramachandran+2006} Ramachandran, R., Demorest, P., Backer, D.~C., Cognard, I., \& Lommen, A.\ 2006, \apj, 645, 303 


\bibitem[Rickett(1990)]{Rickett90} Rickett, B.~J.\ 1990, \araa, 28, 561


\bibitem[Roberts \& Ables(1982)]{ra82}
	Roberts, J.~A., \& Ables, J.~G.  1982,	
    \mnras, 201, 1119
    
    
\bibitem[Romani et al.(1987)]{rbc87}
	Romani, R.~W., Blandford, R.~D., \& Cordes, J.~M.  1987, 
    Nature, 328, 324
    
\bibitem[Savitzky \& Golay(1964)]{SavitzkyGolay} Savitzky, A., \& Golay, M.~J.~E.\ 1964, {\it Anal. Chem.}, 36, 8, 1627-1639

\bibitem[Shannon \& Cordes(2010)]{sc10} Shannon, R.~M., \& Cordes, J.~M.\ 2010, \apj, 725, 1607 

\bibitem[Shannon et al.(2014)]{sod+14} Shannon, R.~M., Os{\l}owski, S., Dai, S., et al.\ 2014, \mnras, 443, 1463 
    
\bibitem[Shannon \& Cordes(2017)]{sc17} Shannon, R.~M., \& Cordes, J.~M.\ 2017, \mnras, 464, 2075 

\bibitem[Spangler et al.(2002)]{Spangler+02} Spangler, S.~R., Kavars, D.~W., Kortenkamp, P.~S., et al.\ 2002, \aap, 384, 654 

\bibitem[Stairs(2002)]{Stairs2002} Stairs, I.~H.\ 2002, Single-Dish Radio Astronomy: Techniques and Applications, 278, 251 

\bibitem[Stinebring et al.(1984)]{Stinebring+84} Stinebring, D.~R., Cordes, J.~M., Rankin, J.~M., Weisberg, J.~M., \& Boriakoff, V.\ 1984, \apjs, 55, 247 

\bibitem[Stinebring et al.(2001)]{Stinebring+01} Stinebring, D.~R., McLaughlin, M.~A., Cordes, J.~M., et al.\ 2001, \apjl, 549, L97 

\bibitem[Taylor(1992)]{Taylor92} Taylor, J.~H.\ 1992, Royal Society of London Philosophical Transactions Series A, 341, 117

\bibitem[Verbiest et al.(2016)]{IPTADR1} Verbiest, J.~P.~W., Lentati, L., Hobbs, G., et al.\ 2016, \mnras, 458, 1267 

\bibitem[You et al.(2007a)]{You+07a} You, X.~P., Hobbs, G., Coles, W.~A., et al.\ 2007, \mnras, 378, 493 

\bibitem[You et al.(2007b)]{You+07b} You, X.~P., Hobbs, G.~B., Coles, W.~A., Manchester, R.~N., \& Han, J.~L.\ 2007, \apj, 671, 907 

\end{thebibliography}
\end{document}